\documentclass[lineno]{jfm}

\usepackage{graphicx}
\usepackage{amsmath}
\usepackage{amssymb}
\usepackage{pifont}
\usepackage{txfonts}
\usepackage{psfrag}
\usepackage[round]{natbib}
\usepackage[T1]{fontenc}
\usepackage{lmodern}
\usepackage{tikz}
\usetikzlibrary{shapes.geometric,shapes.symbols}
\usepackage{scalerel}
\usepackage{upgreek}
\usepackage{booktabs}
\usepackage{verbatim}
\usepackage{framed}
\usepackage{fancyvrb}
\usepackage{varwidth}
\usepackage{bbding}
\usepackage{subfigure}
\usepackage{multirow}
\usepackage[normalem]{ulem}
\usepackage{multicol}
\usepackage{xcolor}
\usepackage{breqn}
\usepackage{soul}
\usepackage{dirtytalk}
\usepackage{upquote}
\usepackage{authblk}
\usepackage{array}
\newcolumntype{P}[1]{>{\centering\arraybackslash}p{#1}}

\usepackage{hyperref}
\hypersetup{
    colorlinks = true,
    urlcolor   = blue,
    citecolor  = black,
}

\newcommand{\RomanNumeralCaps}[1]
\linenumbers

\newcommand{\rahul}[1]{{\color{black}#1}}
\newcommand{\rahuld}[1]{{\color{black}#1}}
\newcommand{\rahulemp}[1]{{\color{black}#1}}

\setstcolor{magenta}
\setul{}{0.5ex}

\newcommand\ufrics{\ensuremath{\overline{U}_{\tau_s}}}
\newcommand\ufrica{\ensuremath{\overline{U}_{\tau_a}}}
\newcommand\ubulk{\ensuremath{\overline{U}_b}}

\newcommand\Reb{\ensuremath{{Re}_b}}

\newcommand\Rets{\ensuremath{{Re}_{\tau_s}}}
\newcommand\Reta{\ensuremath{{Re}_{\tau_a}}}
\newcommand\hmean{\ensuremath{H}}
\newcommand\cylrad{\ensuremath{k}}
\newcommand\cylspacing{\ensuremath{\Lambda}}

\title{Pressure drag reduction via imposition of spanwise wall oscillations on a rough wall}
\author{Rahul Deshpande\aff{1}
  \corresp{\email{raadeshpande@gmail.com}},
  Aman G. Kidanemariam\aff{1},
 and Ivan Marusic\aff{1}}

\affiliation{\aff{1}Dept. Mechanical Engg., University of Melbourne, Parkville, VIC 3010, Australia}

\begin{document}
\maketitle

\begin{abstract}

The present study tests the efficacy of the well-known viscous drag reduction strategy of imposing spanwise wall oscillations to reduce pressure drag contributions in a transitional- and fully-rough turbulent wall flow.
This is achieved by conducting a series of direct numerical simulations of a turbulent flow over two-dimensional (spanwise aligned) semi-cylindrical rods, placed periodically along the streamwise direction with varying streamwise spacing.
Surface oscillations, imposed at \rahul{fixed viscous-scaled} actuation parameters \rahul{optimum} for smooth wall drag reduction, are found to yield substantial drag reduction ($\gtrsim$25\%) for all the rough wall cases, \rahul{maintained at matched roughness Reynolds numbers.}
\rahulemp{While the total drag reduction is due to a drop in both viscous and pressure drag in the case of transitionally-rough flow (i.e. with large inter-rod spacing), it is solely associated with pressure drag reduction for the fully-rough cases (i.e. with small inter-rod spacings), with the latter being reported for the first time.
The study finds that pressure drag reduction in all cases is caused by the attenuation of the vortex shedding activity in the roughness wake, 
in response to wall-oscillation frequencies that are of the same order as the vortex shedding frequencies.}
\rahulemp{Contrary to speculations in the literature, this study confirms that the mechanism behind pressure drag reduction, achieved via imposition of spanwise oscillations, is independent from the viscous drag reduction.}
\rahuld{This mechanism is responsible for weakening of the Reynolds stresses and increase in base pressure in the roughness wake, explaining the pressure drag reduction observed by past studies, across varying roughness heights and geometries.}

\end{abstract}


\section{Introduction and motivation}
\label{intro}

Wall-bounded turbulent flows are common in a wide range of engineering applications, such as flows over ships and submarines, and in pipes. 
In such cases, the bounding walls generally have a `rough' surface topography, which adds a drag penalty in the form of pressure drag.
This drag is generated from the local flow separation and subsequent wake formation behind the protruding roughness elements, \rahul{similar to that noted for bluff bodies \citep{choi2008}}.
An increase in the size of surface roughness increases the \rahul{relative contributions of the pressure drag to the total drag, but reduces the relative} viscous drag contributions \citep{jimenez2004,chung2021}.
The latter is associated with obliteration of the drag-producing near-wall cycle, by roughness, thereby making \rahul{pressure drag the dominant contributor to the total drag in the `fully-rough' regime.} 
These fundamental differences in smooth and rough wall drag-generating mechanisms pose a significant challenge for shipping as well as piping industries, wherein the exposed surfaces degrade from a hydraulically smooth to fully-rough regime in their operation cycle. 
In such cases, a drag reduction strategy that can attenuate both smooth and fully-rough wall drag-generating mechanisms is required, and this forms the focus of the present study.
Here, we will refer to a fully-rough scenario as when the total drag almost entirely comes from the pressure drag.
\rahul{In this respect, it is important to take note of some recent findings \citep{marusic2021} based on high friction Reynolds number ($Re_{\tau}$) experiments, which reveal inertia-dominated scales also make statistically significant contributions to the total skin-friction drag at $Re_{\tau}$ $\gtrsim$ $10^3$.
Contributions from these inertial scales, however, are kept negligible in this study by limiting all our simulations to low $Re_{\tau}$ ($\lesssim$ 300).}

\subsection{\rahul{Literature on rough wall drag reduction}}

Numerous active and passive drag reduction strategies have been proposed in the past \rahul{for turbulent wall-bounded flows} \citep{corke2018}.
However, the majority of these strategies focused on reducing the viscous drag by targeting the near-wall self-sustaining cycle \citep{kim2011,jimenez2018}, rendering them ineffective for attenuating the fully-rough wall drag.
\rahul{Some studies have found large-eddy break-up devices (LEBUs) to be effective in reducing both smooth \citep{corke1982} and rough wall skin-friction drag \citep{bandyopadhyay1986}. 
However, using LEBUs also adds significant parasitic drag, effectively making the total drag reduction negligible.}
Interestingly, recent studies by \citet{banchetti2020}, \citet{nguyen2021} and \citet{garcia2021} found that the well-known strategy of imposing spanwise oscillations on a turbulent rough wall flow can significantly reduce the pressure drag emerging from the roughness.
This is promising since the same drag reduction strategy is also known to substantially reduce the smooth wall (i.e. viscous) drag \citep{akhavan1993,gatti2013,gatti2016,ricco2021}, suggesting continued drag reduction, if implemented on a surface with varying roughness properties over time.
\rahul{Notably, the same strategy has also been found to yield significant drag reduction in the scenario of a transonic flow over a wing \citep{quadrio2022}, indicating success of this strategy across a broad range of flows.}

\rahul{The drag of a rough wall, as well as its overlying boundary layer properties, are a function of the geometric parameters describing the surface topography \citep{chung2021}. 
This has often led to consideration of simplified roughness geometries for a systematic investigation of the influence of these parameters. 
For instance, past studies have considered square \citep{leonardi2003,lee2007} or cylindrical \citep{furuya1976,leonardi2015} rods spanning the entire width of the testing domain (known as two-dimensional or 2-D roughness), or 3-D sinusoidal \citep{chan2015}, hemispherical \citep{wu2020} or cubic roughness \citep{coceal2007,lee2011cube} distributed over the wall.
The same philosophy has also been adopted by past studies understanding the efficacy of drag reduction strategies on the rough wall drag \citep{bandyopadhyay1986,banchetti2020,nguyen2021,garcia2021}.
For example, \citet{banchetti2020} simulated a turbulent channel flow with an isolated 2-D bump on one wall, to generate a pressure drag contribution to the total drag. 
The drag reduction strategy considered by \citet{banchetti2020} involved imposition of streamwise travelling waves of spanwise velocity on the bump wall \citep{quadrio2009}, which was found to enlarge the separation bubble in the bump's wake, while also strongly stabilizing its temporal activity.} 
A pressure drag reduction of $\sim$10\% was noted by \citet{banchetti2020}, \rahul{alongside some viscous drag reduction, for} actuation parameters that previously yielded maximum drag reduction for a smooth wall channel flow.
Similarly, \citet{nguyen2021} simulated a turbulent channel flow with transverse (i.e. 2-D) square bars placed on both walls, positioned at a streamwise offset of 39 times the bar height (imposed by periodicity of the computational domain).
Their study involved imposing a time-oscillating pressure gradient in the spanwise direction across the entire channel cross-section, which yielded a maximum pressure drag reduction of 22\% at an optimum oscillating frequency \rahul{(alongside some viscous drag reduction)}.
Their observations were consistent with \citet{banchetti2020}, who also observed pressure drag reduction due to a decrease in the pressure upstream of the rough element, along with an increase in the base pressure (i.e. downstream of the bar).
\rahul{Based on a control volume analysis of the change in momentum between consecutive roughness elements, \citet{nguyen2021} hypothesized the reduction of the Reynolds shear stresses, in the wake of the roughness elements, to be a manifestation of the reduction of pressure as well as viscous drag.}
\rahuld{A similar weakening of the Reynolds stresses in the roughness wake was noted by \citep{banchetti2020}, suggesting applicability of the same pressure drag reduction mechanism despite differences in the roughness geometry (2-D bump or square roughness).}

\rahulemp{While both of these past studies \citep{banchetti2020,nguyen2021} have significantly improved our understanding, they were confined to a quasi-isolated arrangement of roughness elements, thereby limiting their applicability to transitional rough wall scenarios.
Although no detailed discussion on the pressure drag reduction mechanism was presented in either of these studies, it was speculated to be a direct consequence of the viscous drag reduction.
For instance, \citet{nguyen2021} hypothesized that the pressure drag reduction ``\emph{could be due to the reduced dynamic pressure of the flow encountering the bar}''.
Similarly, \citet{banchetti2020} mentioned ``\emph{the pressure distribution is modified by changes in friction}'', suggesting that pressure drag reduction is dependent on the viscous drag reduction mechanism.
The present study tests this hypothesis by imposing spanwise wall oscillations on a fully rough scenario (i.e. having negligible viscous drag contributions), obtained by maintaining small streamwise offsets between subsequent roughness elements \citep{leonardi2003}.}

\subsection{\rahul{Drawing inspiration from separation control strategies}}
\label{inspire}

Besides being an artefact of the viscous drag-reduced flow upstream of the roughness element, it is plausible that the pressure drag reduction could be a direct consequence of the spanwise oscillations influencing the separating shear layer or roughness wake \citep{choi2008,yakeno2015}.
This motivates some discussion on the findings of previous studies that have investigated the flow physics and control of separated shear layers.
For both streamlined bodies (for example wings at a non-zero angle of attack) and bluff-bodies subjected to separation/pressure drag, one can find several past studies discussing active \citep{seifert2002,post2006,yakeno2015,cho2016,brackston2016} and passive methods \citep{lin1990,son2011} for reducing pressure drag via delay in flow separation.
The control mechanism used in all these studies aims to energize the separated shear layer (or separating boundary layer) to promote flow reattachment, thereby reducing the separation bubble (i.e. wake) size, which subsequently reduces the pressure drag. 
Based on the observations of \citet{banchetti2020} and \citet{nguyen2021}, however, this is not the case when spanwise oscillations are imposed on the flow over 2-D roughness elements (they found separation bubbles to enlarge).
Energizing the boundary layer via momentum injection also enhances the shear layer activity (and consequently the turbulent stresses) in the separated flow, which is also opposite to what is observed in case of rough wall drag reduction via spanwise oscillations \citep{banchetti2020,nguyen2021}.

\rahulemp{Another way to reduce pressure drag, besides reducing the separation bubble size, is by weakening/attenuating the vortex shedding activity in the bluff body wake \citep{choi2008}, which increases its base pressure and consequently reduces pressure drag.
Interestingly, this control mechanism also reduces the Reynolds stresses in the bluff body wake \citep{desai2020,chopra2022}, which matches with observations made by \citet{banchetti2020} and \citet{nguyen2021} in the roughness wake, for drag-reduced cases.}
Several past studies have demonstrated weakening of the shedding activity by enforcing a mismatch in the phase of the separating 2-D shear layer from the bluff body, along its spanwise direction.
For example, \citet{bearman1998} demonstrated suppression of vortex shedding activity by introduction of spanwise waviness on the front stagnation face of a 2-D rectangular bar. 
The same concept was later successfully extended to a circular cylinder by \citet{owen2001}, by changing the straight axis of the cylinder to a sinusoidal one.
\citet{darekar2001} explained that such a 3-D forcing, distorts the quasi 2-D separated shear layers, making them less susceptible to roll up into a Karman vortex street, thereby suppressing the vortex shedding activity.
\rahuld{Their explanation was reinforced later by \citet{hwang2013} who performed linear stability analysis on two-dimensional wakes, thereby revealing that the phenomena is not dependent on the `geometry' of the bluff body.}
The same suppression has also been achieved in a cylinder wake via active control techniques, such as distributed forcing \citep{kim2005}, which imposes blowing and suction on the top and bottom surfaces of the cylinder, with sinusoidally varying intensity along the span.
Considering this background, one would expect imposition of time-varying spanwise oscillations, on the 2-D roughness elements immersed in a wall-bounded flow (which essentially act as bluff bodies), to also disrupt the two-dimensionality of the separating shear layer in the roughness wake \rahuld{(for any roughness geometry)}.

\subsection{\rahul{Present contributions}}

The present study draws inspiration from the previous works of \citet{leonardi2003,leonardi2015} and \citet{nguyen2021} and reports direct numerical simulations of a turbulent flow in an open channel configuration, with 2-D semi-cylindrical rods fixed on the bottom wall in a streamwise periodic manner.
Individual simulations are conducted for different streamwise offsets between rods, thereby enabling investigation of the drag reduction mechanisms for scenarios of high pressure drag domination (small offsets), as well as those with near equal pressure and viscous drag contributions (large offsets).
\rahul{For convenience, we will henceforth refer to the arrangement of semi-cylindrical elements as a `rough' wall, but we note that generalizing the present observations to practical roughness geometries (for example, 3-D, sandpaper, etc.) would require further studies.}
Here, we impose time-periodic wall oscillations in the spanwise direction to achieve drag reduction \citep{akhavan1993}, which is tested here for the first time on fully-rough cases.
\rahulemp{Discussion in $\S$\ref{inspire} suggests that weakening of vortex shedding activity could be the plausible mechanism behind pressure drag reduction noted previously for transitional rough wall cases \citep{banchetti2020,nguyen2021}.
This hypothesis will be tested rigorously in the present study.}

\vspace{-3mm}
\section{Flow configuration and simulation strategy}
\label{setup}

\begin{figure}
   \captionsetup{width=1.0\linewidth}
  \centering

  \begin{minipage}{0.5\textwidth}
    (\textit{a})\\
    \includegraphics[width=\linewidth]{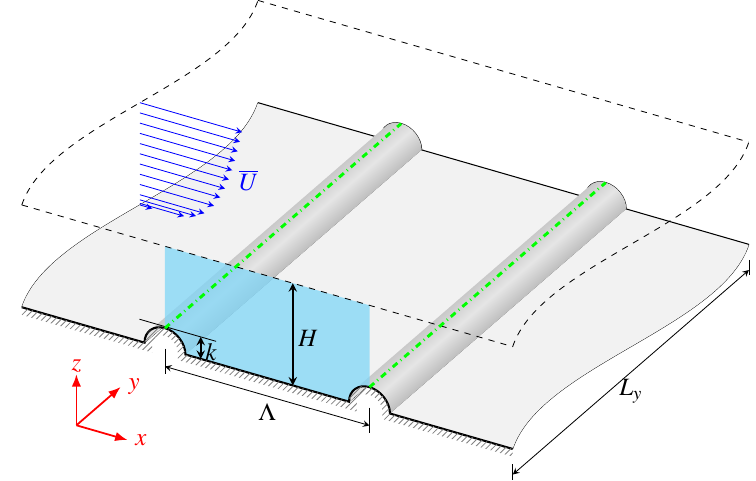}
  \end{minipage}
  \hspace{10pt}
  \begin{minipage}{0.4\textwidth}
    \centerline{(\textit{b})}
    \includegraphics[width=\linewidth]{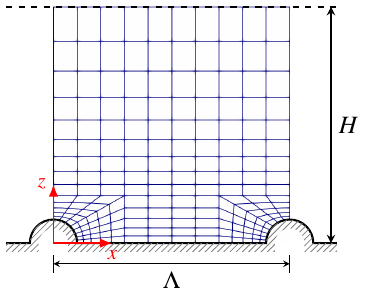}
  \end{minipage}
  \caption{(\textit{a})
    Schematics of an open-channel flow of an incompressible fluid
    over a rough wall, made up of spanwise-aligned semi-cylindrical
    rough elements.
    Terminology is defined in $\S$\ref{setup}.
    The shaded light blue region shows the extent of the domain over
    what is referred to as the streamwise periodic unit (0 $\le$ $x/\Lambda$ $\le$ 1). Dashed green lines are used to indicate the crest of the roughness elements, referred in $\S$\ref{results}.
    (\textit{b})
    Depiction of the two-dimensional quadrilateral spectral elements used to
    discretize the computational domain (shown only over a streamwise periodic unit).}
  \label{fig1}
\end{figure}

As shown in figure \ref{fig1}\textit{a}, we consider an
open-channel flow of an incompressible fluid of density $\rho$ and kinematic
viscosity $\nu$ past a rough bottom wall. The latter is composed of a
series of plane, smooth wall regions parallel to the $x$-$y$ plane, in between semi-cylindrical
rough elements of radius $\cylrad$, that extend across the span and
are spaced at a centre-to-centre distance $\cylspacing$ in the
streamwise direction.  The flow depth \hmean\ is defined as the
wall-normal distance between the bottom plane wall and the upper
free-shear boundary, with roughness height fixed at $k$ = 0.1$H$ \rahuld{for all but one rough wall scenarios ($k$ = 0.2$H$ is considered for one case; see table \ref{tab1})}.  
Throughout this manuscript, we use $x$, $y$ and
$z$ as the streamwise, spanwise and wall-normal directions
respectively, with ($\widetilde{U}$, $\widetilde{V}$, $\widetilde{W}$) representing 
instantaneous velocities in these directions.
\rahulemp{Here, we chose the semi-cylindrical roughness geometry, over the commonly researched square bar geometry \citep{bandyopadhyay1986,leonardi2003,lee2007,nguyen2021}, to avoid a geometrically imposed flow separation location on the individual roughness elements.
While the presence of sharp corners, on the upstream and downstream faces of the square bars, inherently `locks' the flow separation location, the same is not observed when a rounded bar geometry is considered \citep{deshpande2017,alam2022}.
In this way, the choice of a semi-cylindrical roughness element permits an unambiguous investigation of the effect of spanwise wall oscillations on the roughness wake (i.e. pressure drag).}

Two sets of direct numerical simulations of the above configuration
were performed, and are referred to as the static and
actuated cases. In the static cases, the bottom rough wall is kept stationary,
while in the actuated cases the entire bottom wall (including the 
rough elements) oscillates in the spanwise direction at a prescribed oscillation amplitude ($A$) and frequency (1/$T_{osc}$),
expressed as $\widetilde{V}_w\left(t\right)$ = $-A \sin \left( 2{\pi}t/{T_{osc}} \right)$, where $t$ is time.
The no-slip boundary condition is imposed at the bottom wall, whereas
the standard free-slip 
boundary condition mimics the free-surface at the top boundary.
All the simulations are performed in a computational domain that is periodic in
the streamwise and spanwise directions and has a size of   
$L_x \times L_y \times L_z = 6\hmean\times 3\hmean \times \hmean$, 
\rahul{consistent with the domain size considered by previous 
studies investigating flow over 2-D roughness \citep{miyake2001,nagano2004,leonardi2015}.
Notably, \citet{lozano2014} found this box size to be adequate for computing accurate one-point statistics, even in case of a smooth wall channel flow.}
The present simulations are carried out using
the open-source spectral element solver NEKTAR++ \citep{Cantwell2015}.
To exploit
the spanwise homogeneity of the configuration, the domain is discretized by
two-dimensional $N_{elem}$ spectral elements in the $x-z$ plane, combined with
Fourier series expansion in the homogeneous $y$ direction along 192 grid points.
Within each spectral element, velocity and pressure fields are represented by
  Lagrange polynomials of degree 7 \rahul{through the Gauss-Lobatto-Legendre points}, giving viscous-scaled 
  nominal grid resolutions, $\Delta x^+ \approx 1.4$--$5.0$ and $\Delta z^+ \approx 0.6$--$7.3$.
  The spanwise grid resolution is $\Delta{y^+}$ $\approx$ 4.7.
  \rahul{More details of the adopted numerical parameters, and evaluation of the adequacy of the grid
    resolution and domain size, are provided in appendix 1.}  

\begin{table}
   \captionsetup{width=1.0\linewidth}
    \begin{tabular}{llccccccccccc}
      \multicolumn{2}{c}{} & \multicolumn{4}{c}{Static} & \multicolumn{1}{c}{} & \multicolumn{6}{c}{Actuated} \\
      \cmidrule{3-6} \cmidrule{8-13}
      Case & $Re_{b}$ & $\Rets$ & $\overline{C}_{D_{s}}$ & $\overline{C}_{{p}_{s}}$ & $\overline{C}_{{v}_{s}}$ &  & ${T^+_{osc}}$ & $\Reta$ & DR\% & $\overline{C}_{D_{a}}$ & $\overline{C}_{{p}_{a}}$ & $\overline{C}_{{v}_{a}}$  \\
           &         &          & ($\times$10$^{-3}$) & ($\times$10$^{-3}$) & ($\times$10$^{-3}$) & &               &      & & ($\times$10$^{-3}$) & ($\times$10$^{-3}$) & ($\times$10$^{-3}$)\\    
      \noalign{\smallskip}\hline\noalign{\smallskip}
      Smooth &  5045 & 300 &  7.0 &  0.0 & 7.0 & & 100 & 243 & 33.9 & 4.7 & 0.0 & 4.7\\
      $\Lambda$ = 6$k$ & 2323 & 297 & 32.7 & 30.4 & 2.3 & & 100 & 244 & 32.3 & 22.1 & 20.2 & 1.9\\
      $\Lambda$ = 7.5$k$ & 2203 & 299 & 36.6 & 34.4 & 2.2 & & 100 & 242 & 34.7 & 23.9 & 22.4 & 1.5\\
      \multirow{2}{*}{$\Lambda$ = 10$k$} &
      \multirow{2}{*}{2230} & 
      \multirow{2}{*}{301} &
      \multirow{2}{*}{36.3} &
      \multirow{2}{*}{33.1} &
      \multirow{2}{*}{3.2} &
      & 100 & 247 & 32.3 & 24.6 & 23.2 & 1.4\\
      &&&&&&& 10 & 300 & 0.58 & 36.1 & 33.1 & 3.0\\
      \rahuld{$\Lambda$ = 10$k^{*}$} & \rahuld{1617$^{*}$} & \rahuld{299$^{*}$} & \rahuld{68.5$^{*}$} & \rahuld{63.3$^{*}$} & \rahuld{5.3$^{*}$} & & \rahuld{100$^{*}$} & \rahuld{276$^{*}$} & \rahuld{14.7$^{*}$} & \rahuld{58.5$^{*}$} & \rahuld{55.7$^{*}$} & \rahuld{2.8$^{*}$}\\      
      $\Lambda$ = 12$k$ & 2289 & 299 & 33.8 & 29.7 & 4.1 & & 100 & 252 & 27.9 & 24.4 & 22.7 & 1.7\\
      $\Lambda$ = 15$k$ & 2404 & 297 & 30.4 & 25.3 & 5.1 & & 100 & 254 & 26.5 & 22.4 & 20.2 & 2.2\\
      $\Lambda$ = 20$k$ & 2615 & 299 & 26.3 & 19.9 & 6.4 & & 100 & 260 & 25.5 & 19.6 & 16.3 & 3.3\\
      $\Lambda$ = 60$k$ & 3503 & 301 & 14.8 & 6.7 & 8.1 & & 100 & 261 & 24.6 & 11.2 & 5.6 & 5.6
    \end{tabular}
    \caption{A summary of direct numerical simulations for static and actuated cases
      with $A^+$ $=$ $A/{\overline{U}_{{\tau}_{s}}}$ = 12 and oscillation period,
        ${T^+_{osc}}$ $=$ ${T_{osc}}{\overline{U}^2_{{\tau}_{s}}}/{\nu}$.
        \rahuld{Values marked with $^{*}$ correspond to the sole $k$ = 0.2$H$ case, while $k$ = 0.1$H$ for all other cases.}} 
  
  \label{tab1}
\end{table}

The simulation campaign was as follows: initially, \rahuld{eight} static cases \rahul{at varying 6 $\lesssim$ $\Lambda/k$ $\lesssim$ 60} were performed at a target friction Reynolds number, $\Rets$ = ${\ufrics}H/{\nu}$ $\approx$ $300$.
\rahuld{Seven of these eight cases had roughness height $k$ = 0.1$H$ while another had $k$ = 0.2$H$ (for which $\Lambda$ = 10$k$).}
\rahul{Here, $\ufrics$ is the friction velocity calculated based on the momentum balance of the static cases, with $\ufrics^2$ = ${\overline{C}_{{D}_{s}}}{\ubulk^2}/{2}$ (where ${\overline{C}_{{D}_{s}}}$ is the mean total drag coefficient for the static case). 
Detailed information on calculating ${\overline{C}_{{D}_{s}}}$ is given later in this section.}
As a consequence of varying \cylspacing, and in order to ensure a matched $\Rets$, the value of the bulk Reynolds number, $\Reb=\ubulk\hmean/\nu$ varies amongst the different static cases \rahul{(here, $\ubulk \equiv q_f/\hmean$ is the bulk velocity and $q_f$ is the flow rate).
However, this was required to fix the roughness Reynolds number, $k^+$ = ${k}{\ufrics}/{\nu}$ = 30 for all static rough wall cases \rahuld{corresponding to $k$ = 0.1$H$}, since it can dictate the characteristics of the separating shear layer/wake of the roughness element.}
\rahul{It is worth noting here that the present roughness height is a significant fraction of the characteristic outer-scale ($H$), owing to which outer-layer similarity of the velocity statistics can't be expected \citep{jimenez2004}.
However, our $k$ value is smaller than or equal to the roughness height in past studies considering 2-D roughness elements \citep{leonardi2003,leonardi2015,lee2007,nguyen2021}, and has negligible blockage effect on the bulk flow characteristics.}

Next, starting from an established flow field of the static cases, 
their actuated counterparts were simulated such that the same flow rate $q_f$ (and ${Re}_{b}$) 
was maintained for the static and actuated cases at matched $\Lambda/k$.
Here, for all (except one) actuated cases, we consider the optimum wall actuation parameters noted by
\citet{gatti2016} for a smooth wall channel flow, i.e. $A^+ = A/{\overline{U}_{{\tau}_{s}}} = 12$ and
$T^+_{osc} = {T_{osc}}{\overline{U}^2_{{\tau}_{s}}}/{\nu} = 100$.
\rahul{By imposing oscillations at a fixed $T^+_{osc}$ and $A^+$, on rough wall cases having constant roughness Reynolds number ($k^+$ = 30), we can unambiguously investigate effect of spanwise wall oscillations on the roughness wake and pressure drag at various ${\Lambda}/k$ (i.e. from transitional to fully-rough scenarios).}
\rahuld{Majority analysis in this manuscript will be based on $k$ = 0.1$H$ (i.e., $k^+$ = 30) cases, with one case considered for $k$ = 0.2$H$ to demonstrate the generalizability of present conclusions.}
Apart from these rough wall simulations, reference static and
actuated cases with a smooth bottom wall were also run at matched $\Rets$, domain size and oscillation parameter values.
All these simulations have been summarized in table~\ref{tab1}.

\rahul{Considering the imposed streamwise periodicity from roughness elements arranged on the bottom wall, as well as the very low $\Rets$ (which leads to negligible large-scale/inertial contributions), the streamwise extent of the present computational domain ($L_x$ = 6$H$) does not act as a limitation for the majority of the rough wall cases (${\Lambda}$ $\le$ 20$k$). 
In the case of $\Lambda/k$ = 60, however, our numerical domain only had a single roughness element, owing to which flow statistics may not be completely de-correlated from the effects of the computational domain. 
Further, based on the previous investigations of \citet{lozano2014} and \citet{gatti2016}, the present computational domain size may not be sufficient to accurately estimate two-point or higher-ordered  statistics for the smooth wall case.
However, we consider results from these two cases (smooth wall and $\Lambda$ = 60$k$) solely for completeness and so their uncertainties don't influence the conclusions of our study.}

\subsection{\rahul{Definition of flow statistics}}

For the purpose of analyzing flow properties in this manuscript, we 
define a spanwise and time-averaging operator $\langle \varphi \rangle$
of any instantaneous flow quantity $\widetilde{\varphi}$, that
exploits the inherent periodicity of length \cylspacing\  along $x$ (figure \ref{fig1}a) following:
\begin{equation}\label{eq:operator-1}
  \langle \varphi \rangle (x,z) =
  \frac{1}{N_{\rm cyl}}\frac{1}{T_{\rm obs}}\frac{1}{L_y}
  \sum_{i=1}^{N_{\rm cyl}}
  \int^{\rahul{T_{\rm obs}}}_{0} \int^{L_{y}}_{0}
  \widetilde{\varphi} \;{\rm d}y{\rm d}t\;,
\end{equation}
where $T_{\rm obs}$ is the simulation steady-state interval and
$N_{\rm cyl} =L_x/\cylspacing$ is the number of rough elements or streamwise periodic units. 
In \eqref{eq:operator-1}, $x$ spans from the centre of one rough element 
($x=0$) to that of the next downstream ($x=\Lambda$), which we refer to here 
as the streamwise periodic unit of the computational domain (0 $\le$ $x/\Lambda$ $\le$ 1).
Figure \ref{fig1}(a) highlights this periodic unit in light blue shading for reference.

In the presence of the rough elements, the $x$-component of the instantaneous
hydrodynamic force acting upon the bottom wall (i.e. the \rahul{total} drag, $\widetilde{D}$), 
comprises both viscous ($\widetilde{V}$) and pressure ($\widetilde{P}$) force contributions. 
The streamwise variation of the time- and span-averaged \rahul{total} drag force
(per unit streamwise width) is given as:
\begin{equation}\label{eq:hydrodynamic-force-1}
  {\langle{D}\rangle}(x) =
  \langle \sigma_{1j}n_j\rangle{L_y}
  = \underbrace{\langle \tau_{1j}n_j \rangle{L_y}}_{{\langle{V}\rangle}(x)} \;
  \underbrace{- \langle {p\delta_{1j}n_j} \rangle{L_y}}_{{\langle{P}\rangle}(x)}, 
\end{equation}
where $\boldsymbol{n}$ is  the unit normal and $p$ is the hydrostatic pressure (i.e. different from pressure force, $P$).
We can further integrate \eqref{eq:hydrodynamic-force-1} in the streamwise direction
to obtain the total hydrodynamic force (i.e. the \rahul{total} drag, $\overline{D}$) acting on the rough wall as:
\begin{equation}\label{eq:hydrodynamic-force-2}
  \overline{D} =
  \int^{\Lambda}_{0} {\langle{\boldsymbol{D}}\rangle}(x) {\;}{\rm d}S
  \;=\; \underbrace{\int^{\Lambda}_{0}
    {{\langle{V}\rangle}(x)}{\;}{\rm d}S}_{\overline{V}} \;
  + \underbrace{\int^{\Lambda}_{0} {{\langle{P}\rangle}(x)}{\;}{\rm d}S}_{\overline{P}},
\end{equation}
where ${\rm d}S$ is the differential surface length along the bottom boundary.
Note that the pressure drag force, ${\langle}{P}{\rangle}$($x$) = 0 over the plane sections in between the semi-cylindrical rough elements (figure \ref{fig1}a), while $\overline{V}$ is the sum of viscous force contributions from both the plane and semi-cylindrical sections.
Following (\ref{eq:hydrodynamic-force-2}), the \rahul{total} drag coefficient,
$\overline{C}_{D}$ = $2\overline{D}/(\rho{\ubulk^2}\cylspacing L_y)$ =  $\overline{C}_{p} + \overline{C}_{v}$, 
where $\overline{C}_{p} = 2\overline{P}/(\rho{\ubulk^2}{\cylspacing L_y})$
and $\overline{C}_{v} =2\overline{V}/(\rho{\ubulk^2}{\cylspacing L_y})$
are respectively the coefficients of pressure and viscous force contributions to $\overline{C}_{D}$.
Similarly, the streamwise-resolved viscous force coefficient,  
$\langle C_{v} \rangle(x) = 2\langle V \rangle/(\rho {\ubulk^2} L_y)$.
To differentiate the force coefficients associated with the static and actuated cases, subscripts `$s$' and {`$a$'} are used respectively, which are tabulated in table \ref{tab1} for all the simulation cases.
Here, with $Re_{b}$ kept constant, the friction velocities are different for the static ($\ufrics$) and actuated ($\ufrica$) cases, which are determined directly from the mean streamwise momentum balance by following:
$\ufrics^2$ = ${\overline{C}_{{D}_{s}}}{\ubulk^2}/{2}$ and 
$\ufrica^2$ = ${\overline{C}_{{D}_{a}}}{\ubulk^2}/{2}$.
This yields the definition of the \rahul{total} drag reduction, which is:
DR\% = ${100{\times}}({{\overline{C}_{{D}_{s}}} - {\overline{C}_{{D}_{a}}}})/{\overline{C}_{{D}_{s}}}$ = ${100{\times}}({ {\ufrics^2} - {\ufrica^2} })/{{\ufrics^2}}$,
\rahul{reported in table \ref{tab1}.}
\rahul{For example,} DR\% for the present smooth wall case is 33.9\% which, \rahul{despite the marginal computational box size \citep{gatti2016},} lies within the predictions made by \citet{gatti2013,gatti2016} at this $\Rets$.

\vspace{-3mm}
\section{Results and discussions}
\label{results}

\begin{figure}
   \captionsetup{width=1.0\linewidth}
\centering
\includegraphics[width=1.0\textwidth]{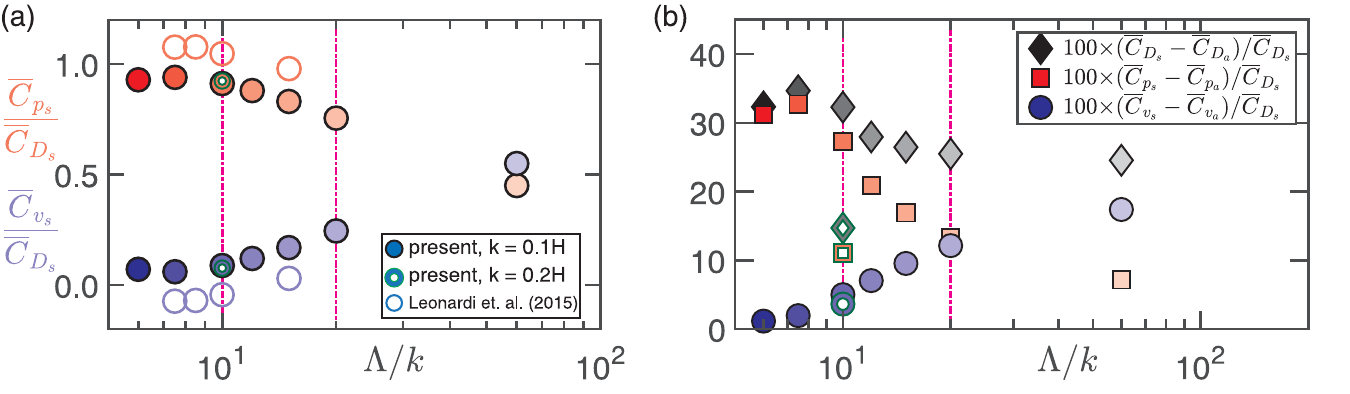}
\caption{(a) Fractional contribution to the total drag experienced by a static rough wall (${\overline{C}_{D_{s}}}$), from viscous forces (${\overline{C}_{v_{s}}}$; in blue shading) and pressure forces (${\overline{C}_{p_{s}}}$; in red shading), for cases of varying streamwise offsets between rough elements ($\Lambda/k$).
(b) \rahul{Total} percentage drag reduction (in black diamonds), percent drag reduction due to decrease in pressure drag (in red squares), and due to decrease in viscous drag (in blue circles), on imposition of wall oscillations at $T^+_{osc}$ = 100.
\rahuld{In (a,b), fully and partially filled symbols correspond to present simulations with $k$ = 0.1$H$ and 0.2$H$ respectively, while empty symbols in (a) correspond to data of \citet{leonardi2015}.}
Dash-dotted magenta lines highlight cases: $\Lambda$ = 10$k$ and 20$k$ analyzed in figures \ref{fig3}-\ref{fig5}.}
\label{fig2}
\end{figure}

\begin{figure}
   \captionsetup{width=1.0\linewidth}
\centering
\includegraphics[width=1.0\textwidth]{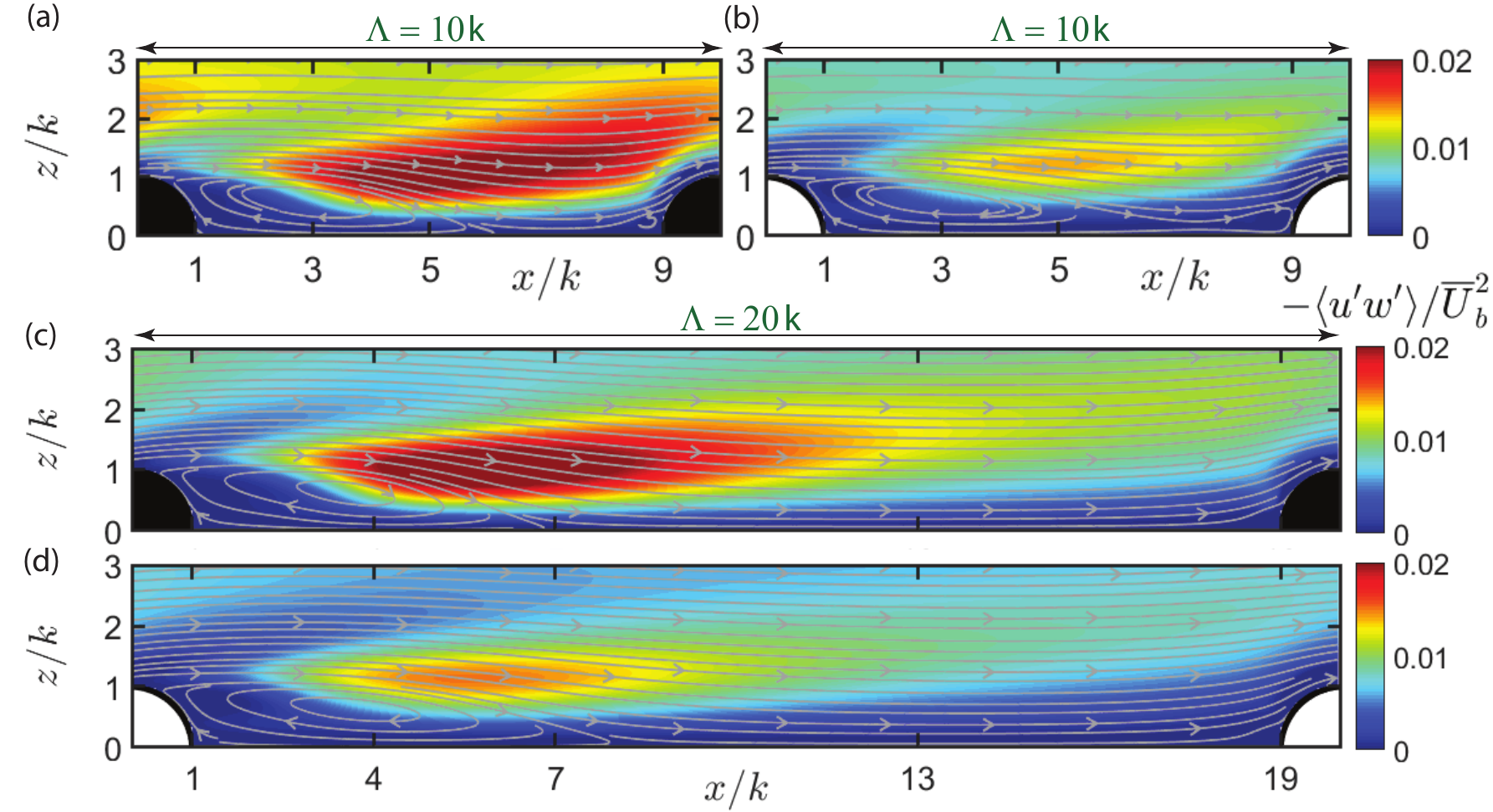}
\caption{\rahul{Non-dimensional Reynolds shear stresses, $-{\langle}{{u'}{w'}}{\rangle}$/$\overline{U}^2_b$, (in coloured contours)
  and mean flow streamlines (represented by solid grey lines) estimated for the streamwise periodic unit of the rough wall simulations (0 $\le$ $x/\Lambda$ $\le$ 1), in the wall-normal plane.
    (a) $\Lambda = 10k$, static;
    (b) $\Lambda = 10k$, actuated at $T^+_{osc} = 100$;
    (c) $\Lambda = 20k$, static;
    (d) $\Lambda = 20k$, actuated at $T^+_{osc} = 100$.
    \rahuld{All plots correspond to the simulations with roughness height, $k$ = 0.1$H$.}}} 
\label{fig3}
\end{figure}

\begin{figure}
   \captionsetup{width=1.0\linewidth}
\centering
\includegraphics[width=1.0\textwidth]{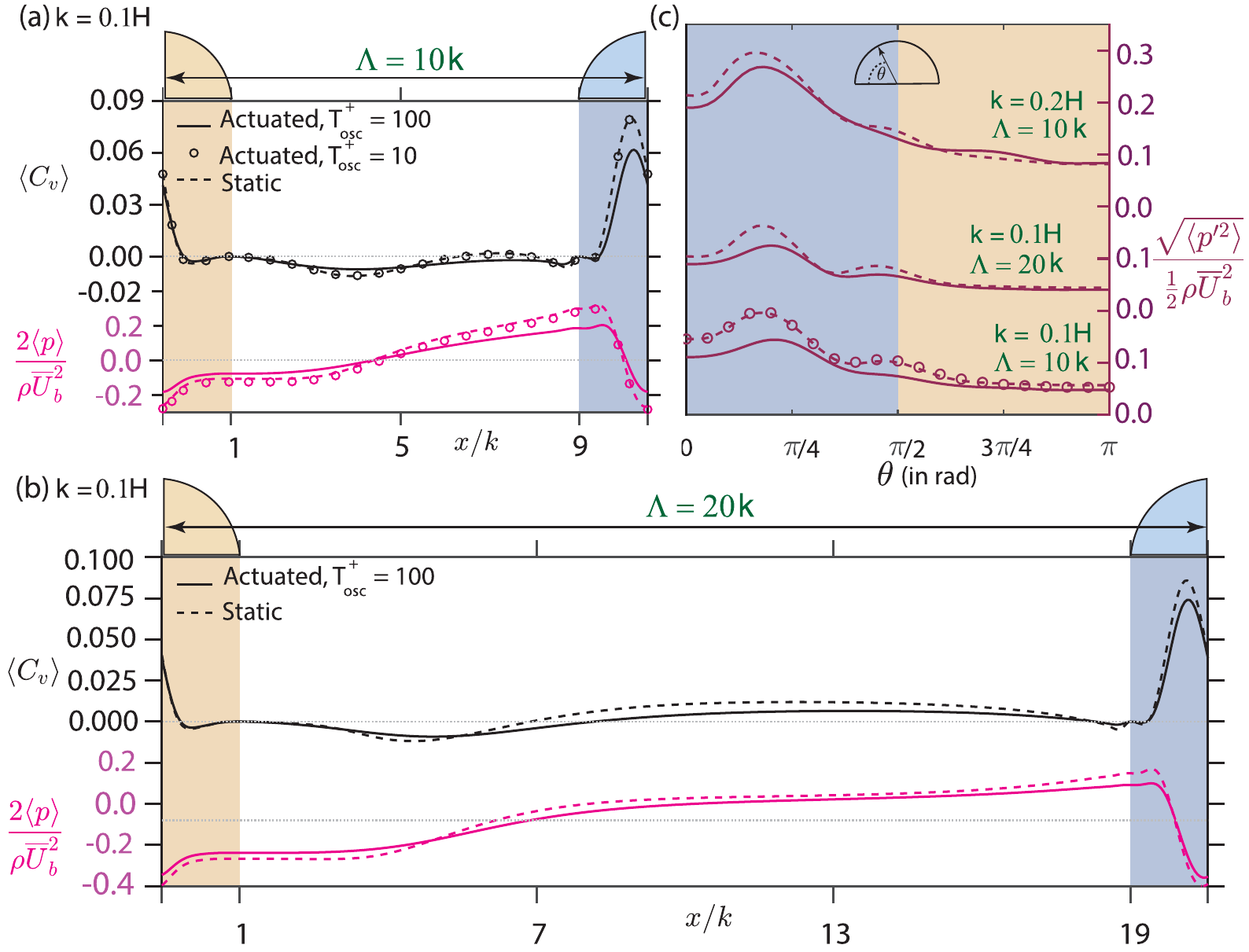}
\caption{\rahul{(a,b)} Span- and time-averaged viscous force coefficients (${\langle}{C_v}{\rangle}$; in black) and normalized surface-pressure ($2\langle p \rangle/\rho\ubulk^2$; in magenta), plotted as a function of $x/k$ in the streamwise periodic unit (0 $\le$ $x/\Lambda$ $\le$ 1), for the \rahuld{$k$ = 0.1$H$} rough wall cases: (a) $\Lambda$ = 10$k$ and (b) $\Lambda$ = 20$k$.
(c) Root-mean-square of the surface-pressure fluctuations (${ 2\sqrt{\langle {p'}^2 \rangle}/\rho\ubulk^2 }$) along the curvature (0 $\le$ $\theta$ $\le$ $\pi$) of the semi-cylindrical rough elements \rahuld{for the same two cases of $k$ = 0.1$H$ as in (a,b) and one of $k$ = 0.2$H$}.
In (a-c), blue and yellow shading represents upstream (0 $\le$ $\theta$ $\le$ $\pi$/2) and downstream ($\pi$/2 $\le$ $\theta$ $\le$ $\pi$) sections of the semi-cylindrical rough element.}
\label{fig4}
\end{figure}

We begin by comparing the present static rough wall statistics with previously published data \citep{leonardi2015}.
For this purpose, results presented in figures \ref{fig2}(a), \ref{fig3}(a,c) and \ref{fig4}(a,b) will be discussed first, before proceeding towards investigating the change in statistics owing to drag reduction.
Figure \ref{fig2}(a) shows fractions of the \rahul{total} drag coefficient ($\overline{C}_{{D}_{s}}$), coming from the pressure drag ($\overline{C}_{{p}_{s}}$) and viscous drag ($\overline{C}_{{v}_{s}}$) contributions, \rahuld{for cases corresponding to both $k$ = 0.1$H$ and 0.2$H$}.
For cases of $\Lambda/k$ $\lesssim$ 10, $\overline{C}_{{p}_{s}}$ is essentially the sole contributor to $\overline{C}_{{D}_{s}}$, with relative contributions from $\overline{C}_{{v}_{s}}$ gradually increasing (and those from $\overline{C}_{{p}_{s}}$ decreasing) with increasing $\Lambda/k$ $\gtrsim$ 10.
At $\Lambda/k$ $=$ 60, \rahul{notably}, both contribute nearly equally to $\overline{C}_{{D}_{s}}$.
\rahuld{Further, for $\Lambda/k$ $=$ 10, the pressure and viscous contributions to the total drag are found to be similar irrespective of $k$ = 0.1$H$ or 0.2$H$.}
This nature of variation of $\overline{C}_{{p}_{s}}$ and $\overline{C}_{{v}_{s}}$, with $\Lambda$/$k$, is consistent with \citet{leonardi2015}, suggesting the mean flow physics observed by the latter can be extended to the present study. 
This is despite \citet{leonardi2015} employing a circular rod geometry for the rough elements, \rahuld{and conducting simulations for different ${\Lambda}/k$ at a constant bulk flow Reynolds number, both of which are different from the approach adopted in the present study.
\citet{leonardi2015} also noted that the variation of pressure and viscous drag with ${\Lambda}/k$ did not change significantly on increasing the flow Reynolds number (at least in the low Reynolds number regime).}
\rahuld{This establishes that a similar variation in pressure drag domination can be expected on changing the streamwise offsets between 2-D rough elements (for 0.1$H$ $\le$ $k$ $\le$ 0.2$H$), irrespective of differences in roughness geometry, flow Reynolds number and simulation approach (i.e. constant friction versus bulk Reynolds numbers).
Figure \ref{fig2}(a), hence, demonstrates that} the present rough wall cases are valid for testing the efficacy of spanwise wall oscillations on varying degrees of pressure drag domination.
\rahul{Based on our definition given in $\S$\ref{intro}, we henceforth refer to the $\Lambda/k$ $\lesssim$ 10 cases as fully-rough, while those corresponding to 10 $<$ $\Lambda/k$ $\le$ 60 referred to as transitionally rough.}

The flow physics associated with the variation of $\overline{C}_{{v}_{s}}$ and $\overline{C}_{{p}_{s}}$ can be understood based on figures \ref{fig3}(a,c) and \ref{fig4}(a,b). 
\rahul{The former depicts the normalized Reynolds shear stresses ($\langle{u'}{w'}\rangle$/${\ubulk}^2$) and mean flow streamlines, while the latter shows} the time- and span-averaged viscous force coefficients (${\langle}{C_v}{\rangle}$) and normalized surface-pressure ($2\langle p \rangle/\rho\ubulk^2$), plotted across the streamwise periodic unit (0 $\le$ $x/{\Lambda}$ $\le$ 1) for $\Lambda$ = 10$k$ and 20$k$ \rahuld{with $k$ = 0.1$H$}.
These two cases were selected \rahul{as representatives of the fully and transitionally rough scenarios, respectively}.
\rahul{Here, $u'$ and $w'$ are respectively the streamwise and wall-normal velocity fluctuations obtained via $u'$ $=$ $\widetilde{U}$ $-$ $\langle U \rangle$.}
It is evident from figure \ref{fig4}(a) that ${\langle}{C_v}{\rangle}$ $\lesssim$ 0 across the plane region between the rough elements (1 $\le$ $x/k$ $\le$ 9), for $\Lambda$ = 10$k$.
On the other hand, for $\Lambda$ = 20$k$, ${\langle}{C_v}{\rangle}$ $<$ 0 only immediately downstream of the rough element (1 $\le$ $x/k$ $\le$ 9), after which ${\langle}{C_v}{\rangle}$ $>$ 0 for 7 $\le$ $x/k$ $\le$ 19 (figure \ref{fig4}b).
These \rahul{trends are consistent with observations from the mean flow field in figure \ref{fig3}(a), wherein the wake/re-circulation region can be noted across the majority of the plane region between subsequent} rough elements, in case of small streamwise offsets ($\Lambda$ $\lesssim$ 10$k$).
\rahuld{Qualitatively similar mean flow fields and surface statistics (${\langle}{C_v}{\rangle}$, ${\langle}p{\rangle}$) were noted in case of $k$ = 0.1$H$ and 0.2$H$ for $\Lambda$ = 10$k$ (not shown here for brevity).}
For larger \rahul{$\Lambda/k$ (figure \ref{fig3}c)}, however, the separated shear layer from upstream roughness reattaches in the plane region between these elements \rahul{($x$ $\sim$ 7$k$)}, leading to \rahul{an extended region of} wall-attached flow impinging onto the downstream roughness element (resulting in ${\langle}{C_v}{\rangle}$ $>$ 0). 
While the variation of ${\langle}{C_v}{\rangle}$($x$) \rahul{in the plane region is opposite for the transitional and fully rough cases (figure \ref{fig4})}, the nature of variation of the surface-pressure is very similar, with  $\langle p \rangle$($x$) much higher on the upstream (relative to the downstream) section of the rough element, which generates the pressure drag.
\rahul{Notably, both of the present results plotted in figures \ref{fig3}(a,c) and \ref{fig4}(a,b) are consistent with the \rahuld{observations made previously by \citet{leonardi2015}, explaining the similar trends in pressure and viscous drag noted in figure \ref{fig2}(a).}}

With the static cases \rahul{demonstrated to be consistent with the literature}, we next investigate the drag reduction achieved on imposing spanwise wall oscillations for the same rough wall cases.
Figure \ref{fig2}(b) plots DR\%, on imposition of wall-oscillations at $A^+$ $=$ 12 and $T^+_{osc}$ $=$ 100, for various ${\Lambda}/k$.
The plot suggests substantial DR\% is achieved for all wall-oscillation cases at $T^+_{osc}$ $=$ 100 \rahuld{corresponding to $k$ = 0.1$H$.
The DR\% associated with $k$ = 0.2$H$, however, is not as high as noted for $k$ = 0.1$H$ at $\Lambda$ = 10$k$, and this difference is likely owing to the same spanwise forcing magnitude ($A^+$ = 12) imposed on flows with significantly different pressure drag magnitudes (nearly double for $k$ = 0.2$H$ than 0.1$H$; see table \ref{tab1}).}
\rahulemp{The most interesting observation is the maximum DR ($\approx$ 34\%) noted at ${\Lambda}/k$ = 7.5, which is at par to that noted via pure viscous drag reduction for a smooth wall (for the same $A^+$ and $T^+_{osc}$).
This is a novel result and is particularly notable considering our findings in figure \ref{fig2}(a), which indicate negligible viscous drag contributions at ${\Lambda}/k$ = 7.5.
It suggests this DR\% is solely from pressure drag reduction, which has to be driven by an independent mechanism.}
To quantify what fraction of the \rahul{total} DR\% is achieved by decreasing the pressure and viscous drag, figure \ref{fig2}(b) also shows 100\:${\times}$\:($\overline{C}_{{p}_{s}}$ - $\overline{C}_{{p}_{a}}$)/$\overline{C}_{{D}_{s}}$ and 100\:${\times}$\:($\overline{C}_{{v}_{s}}$ - $\overline{C}_{{v}_{a}}$)/$\overline{C}_{{D}_{s}}$ for varying $\Lambda/k$.
For significantly large streamwise offsets (10$k$ $<$ $\Lambda$ $\le$ 60$k$), which is the scenario investigated by \citet{banchetti2020} and \citet{nguyen2021}, both viscous drag reduction and pressure drag reduction contribute to the \rahul{total} DR\% significantly.
This can be understood by comparing \rahul{the mean flow field (figures \ref{fig3}c,d), as well as} ${\langle}{C_v}{\rangle}$($x$) and $\langle p \rangle$($x$) between static and actuated cases (figure \ref{fig4}b), for $\Lambda$ = 20$k$.
\rahul{Consistent with \citet{banchetti2020}, the pressure drag reduction can be associated with extension of the wake size downstream of the roughness elements (refer to streamlines in figures \ref{fig3}c,d), while viscous drag reduction can be associated with the weakening of the near-wall turbulence in the wall attached flow between the elements (refer to Reynolds shear stresses at $z$ $<$ 0.1H).
This is supported quantitatively by figure \ref{fig4}(b), where} ${\langle}{C_v}{\rangle}$ immediately upstream of the rough element (7 $\lesssim$ $x/k$ $\lesssim$ 19) reduces, compared to the static case.
This is accompanied by reduction in ${\langle}{C_v}{\rangle}$ in the upstream section of the semi-cylindrical surface, combination of which leads to a considerable viscous drag reduction noted in figure \ref{fig2}(b).
Similarly, $2\langle p \rangle/\rho\ubulk^2$ can be noted to decrease in the upstream section of the rough element (7 $\lesssim$ $x/k$ $\lesssim$ 19.5; figure \ref{fig4}b), while it increases downstream of the roughness (0 $\lesssim$ $x/k$ $\lesssim$ 4).
This explains the drop in pressure drag in figure \ref{fig2}(b), which \citet{nguyen2021} had associated with the decrease in the dynamic pressure in 7 $\lesssim$ $x/k$ $\lesssim$ 19.
\rahul{Consistent with the literature \citep{banchetti2020,nguyen2021}, the drag reduced flow for $\Lambda$ = 20$k$ also corresponds with weakening of the Reynolds shear stresses in the roughness wake ($z$ $\approx$ 0.1H; figures \ref{fig3}c,d).}
\rahuld{This reaffirms the weakening of the Reynolds shear stresses on pressure drag reduction, irrespective of differences in the roughness geometries (semi-cylindrical, cylindrical or square)}.

As ${\Lambda}/k$ decreases along 60 $\ge$ ${\Lambda}/k$ $\ge$ 10, the streamwise extent of the wall attached flow between rough elements reduces (figures \ref{fig3}a,c), which reduces the contribution of viscous drag reduction to the \rahul{total} DR\% (figure \ref{fig2}b).
For 5 $\lesssim$ $\Lambda$/$k$ $\lesssim$ 10, owing to the absence of flow reattachment/wall-attached flow between rough elements  (i.e. ${\langle}{C_v}{\rangle}$($x$) $\lesssim$ 0; \citealp{leonardi2015}), almost all the drag reduction is associated with a decrease in the pressure drag.
\rahuld{This is observed irrespective of the variation in roughness height, i.e. for both $k$ = 0.1$H$ and 0.2$H$.
The dominant pressure drag reduction} can be confirmed by comparing ${\langle}{C_v}{\rangle}$ and ${\langle}{p}{\rangle}$ plotted for the static and actuated cases ($T^+_{osc}$ = 100) at $\Lambda$ = 10$k$ and $k$ = 0.1$H$ (figure \ref{fig4}a), where one can note the drop in ${\langle}{p}{\rangle}$ upstream of the roughness (5 $\lesssim$ $x/k$ $\lesssim$ 9.5), and increase in ${\langle}{p}{\rangle}$ downstream of the roughness (0 $\lesssim$ $x/k$ $\lesssim$ 4), while there is negligible variation in ${\langle}{C_v}{\rangle}$ across the plane region (1 $\lesssim$ $x/k$ $\lesssim$ 9).
The same can be understood qualitatively by comparing the mean flow fields of the static and actuated cases depicted for $\Lambda$ = 10$k$, in figures \ref{fig3}(a,b).
In the static scenario, a very small region of wall-attached mean flow can be noted between the two roughness elements, which however completely vanishes on imposition of wall oscillations.
\rahulemp{The fact that significant pressure drag reduction is noted without any changes in the viscous drag (for 6 $\lesssim$ $\Lambda$ $\lesssim$ 10$k$) suggests existence of a unique pressure drag reduction mechanism.}
\rahuld{This observation is valid for both $k$ = 0.1$H$ and 0.2$H$ (figure \ref{fig2}b), and confirms that the same pressure drag reduction mechanism would be applicable for varying roughness heights.}
\rahulemp{These results are novel, and confirms that pressure drag reduction in a 2-D rough wall is not dependent on the viscous drag reduction.
Another new observation from figures \ref{fig3}(a,b) is that the Reynolds shear stresses are attenuated in the wake of the roughness elements, even when viscous drag reduction is negligible.}
It suggests that this attenuation of the turbulent stresses, which are noted across 6 $\le$ $\Lambda/k$ $\le$ 60 (not shown), is likely associated with the pressure drag reduction mechanism.
\rahuld{Despite differences in their roughness geometries compared to the present study, both \citet{banchetti2020} and \citet{nguyen2021} noted a similar reduction in the Reynolds stresses and an increase in base pressure in their roughness wake (on imposition of oscillations).
These consistent results suggest that the same pressure drag reduction mechanism is applicable for different roughness geometries, which, however can only be confirmed after conducting dedicated simulations (which is beyond the present scope).}

\subsection{\rahul{Mechanism behind pressure drag reduction}}

Referring to our previous discussion in $\S$\ref{inspire}, regarding \rahul{bluff body} drag reduction based on increases in base pressure (i.e. ${\langle}{p}{\rangle}$ at $x/k$ $\sim$ 1) \rahul{accompanied by weakening of Reynolds stresses}, our observations from figures \ref{fig3} and \ref{fig4}(a,b) point towards weakening of the vortex shedding in roughness wake as the plausible pressure drag reduction mechanism \citep{choi2008}.
This can be investigated by analyzing the surface-pressure fluctuations (${p'}$) on the rough element; past studies on the flow past a circular cylinder \citep{deshpande2017,desai2020,chopra2022} have shown that weakening of its shedding activity is reflected by a drop in the lift force fluctuations, as well as surface-pressure fluctuations at the crest of the cylinder (i.e. at $\theta$ = $\pi/2$, where $\theta$ is the azimuthal angle from the front stagnation point).
This sensitivity is owing to its proximity to the azimuthal location, from where the shear layer separates and sheds into the wake (0.4$\pi$ $\le$ $\theta$ $\le$ 0.65$\pi$; \citealp{desai2020,chopra2022}).
With this background, we consider ${p'} (x,y,z,t) = {\widetilde{p}} (x,y,z,t) -  \langle {p} \rangle (x)$ over the surface of the cylinder, where tilde (\:${\widetilde{\;}}$\:) and prime (\:$'$\:) respectively denote instantaneous and fluctuating properties.
Figure \ref{fig4}(c) plots the normalized root-mean-square of the surface-pressure fluctuations (${ 2\sqrt{\langle {p'}^2 \rangle}/\rho\ubulk^2 }$) across the surface of the semi-cylindrical rough element, with 0 $\le$ $\theta$ $\le$ $\pi$/2 in blue and $\pi$/2 $\le$ $\theta$ $\le$ $\pi$ in yellow shading.
It is evident that pressure fluctuations drop at $\theta$ $=$ $\pi$/2, on imposition of wall oscillations at $T^+_{osc}$ = 100, for both $\Lambda$ = 10$k$ and 20$k$ \rahuld{and different roughness heights ($k$ = 0.1$H$ and 0.2$H$).
Notably, the drop in pressure fluctuations for $k$ = 0.2$H$ is relatively smaller than that for $k$ = 0.1$H$ at $\theta$ $=$ $\pi$/2, and this is consistent with the relatively lower DR\% noted in case of the former (figure \ref{fig2}b).}

\begin{figure}
   \captionsetup{width=1.0\linewidth}
\centering
\includegraphics[width=1.0\textwidth]{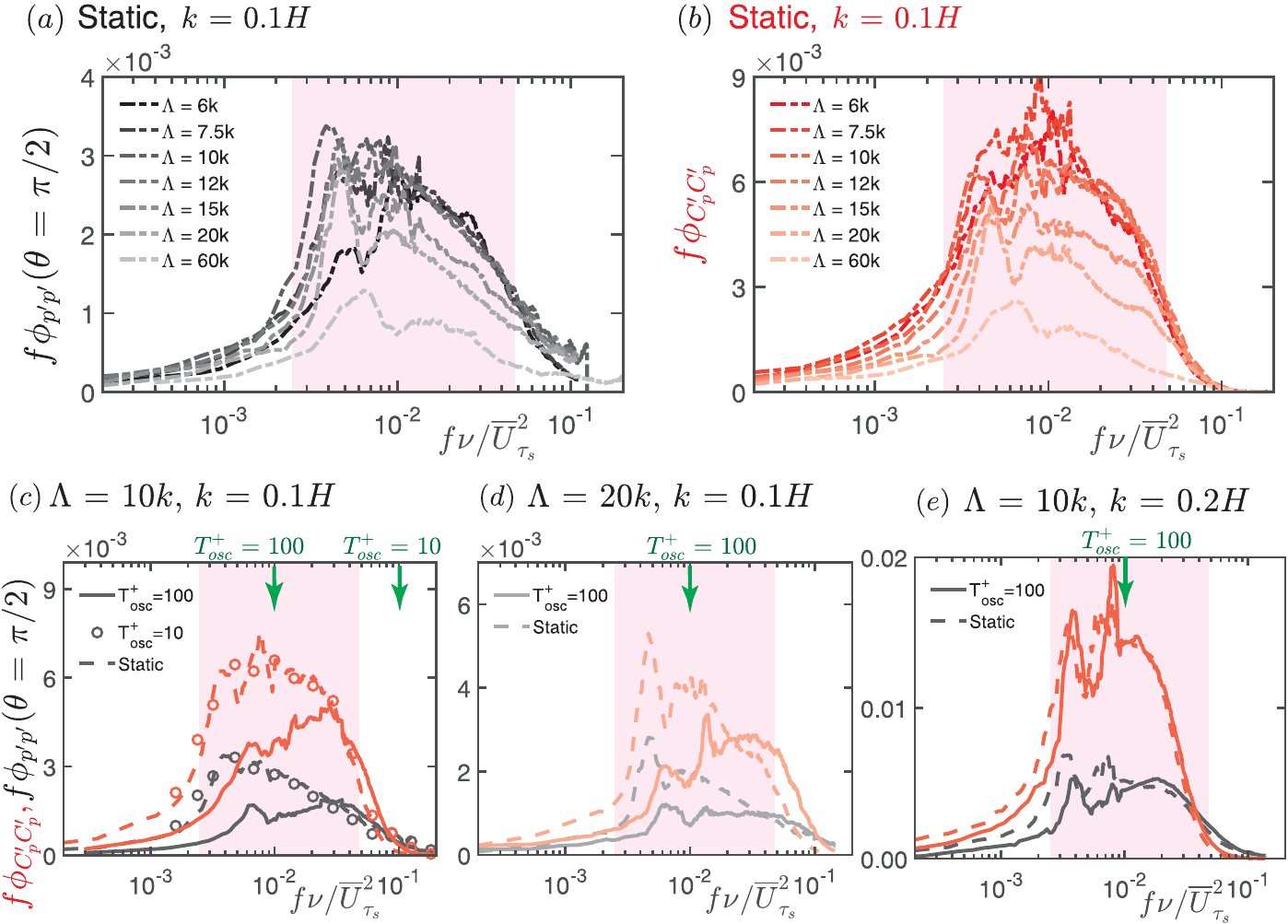}
\caption{\rahul{Premultiplied frequency spectra of the fluctuating surface-pressure (${f}{{\phi}_{{p'}{p'}}}$; in black) at the crest of the rough element ($\theta$ = $\pi$/2), and of the fluctuating pressure force coefficient (${f}{{\phi}_{{C'_p}{C'_p}}}$; in red), as a function of the viscous-scaled frequency ($f{\nu}/{\overline{U}^{2}_{{\tau}_{s}}}$) for various rough wall scenarios. 
(a) and (b) respectively plot ${f}{{\phi}_{{p'}{p'}}}$ and ${f}{{\phi}_{{C'_p}{C'_p}}}$ for the static rough wall cases at 6 $\le$ ${\Lambda}/k$ $\le$ 60.
\rahuld{(c-e) compare both the spectra for the static and actuated cases at (c) $\Lambda$ = 10$k$, $k$ = 0.1$H$, (d) $\Lambda$ = 20$k$, $k$ = 0.1$H$, and (e) $\Lambda$ = 10$k$, $k$ = 0.2$H$.}
Green arrows are used to indicate (${T_{osc}}{\overline{U}^{2}_{{\tau}_{s}}}/{\nu}$) on the $x$-axis, and the magenta shaded background nominally represents the most energetic frequency range for both the spectra (10$^{-2}$ $\lesssim$ $f{\nu}/{\overline{U}^{2}_{{\tau}_{s}}}$ $\lesssim$ 10$^{-1}$). 
Here, ${f}{{\phi}_{{p'}{p'}}}$ is normalized by (${\rho}{\overline{U}^2_{b}}$/2)$^2$.}}
\label{fig5}
\end{figure}

The fact that wall-actuations weaken the vortex shedding suggests that the actuation frequency (1/$T^+_{osc}$) must have a connection with the shedding frequency.
Hence, we test the sensitivity of pressure drag reduction to $T^+_{osc}$, by imposing actuations differing by an order of magnitude.
Interestingly, $T^+_{osc}$ = 10 yields negligible pressure and \rahul{total} drag reduction for $\Lambda$ $=$ 10$k$ (table \ref{tab1}).
Indeed, when ${\langle}{p}{\rangle}$, ${\langle}{C_v}{\rangle}$ and ${ \sqrt{\langle {p'}^2 \rangle} }$ profiles are plotted for the $T^+_{osc}$ = 10 actuation case in figure \ref{fig4}(a), the corresponding profiles overlap with those of the static case, suggesting negligible changes in the mean flow physics.
\rahulemp{We can thus conclude that pressure drag reduction, on imposition of spanwise wall oscillations, is associated with weakening of the vortex shedding activity, and this explains the sensitivity of DR\% to $T^+_{osc}$ (also see \citealp{nguyen2021}).}
Further information for interested readers, this weakening of the vortex shedding activity has also been confirmed by comparing the mean spanwise vorticity in the roughness wake, between static and actuated cases, but is not shown here for brevity.

In order to confirm the causal relationship that weakening of shedding activity leads to pressure drag reduction, we investigate the spectral distribution of $p'$ at $\theta$ = $\pi/2$, and of the fluctuating pressure-force coefficient ($C'_p$). 
Here, $C'_p$ is obtained via: $C'_p$($y$,$t$) = $\int^{L_x}_{0}$ ${\widetilde{C_p}}(x,y,t) {\;}{\rm d}S$ $-$ $\overline{C}_p$, where d$S$ is the differential streamwise distance across the entire bottom wall.
We analyze their energy spectra in the time/frequency domain to relate the spectral distribution to the imposed wall-oscillation frequency ($1/{T^{+}_{osc}}$).
Figures \ref{fig5}(a,b) respectively plot the two spectra in the premultiplied form (i.e. ${f}{{\phi}_{{p'}{p'}}}$ and ${f}{{\phi}_{{C'_p}{C'_p}}}$), for all the static rough wall cases (6 $\le$ ${\Lambda}/k$ $\le$ 60) \rahuld{corresponding to $k$ = 0.1$H$}.
They are plotted against the frequency scale ($f$), \rahul{which is normalized in viscous units ($f{\nu}/{\overline{U}^{2}_{{\tau}_{s}}}$).}
Here, considering that the semi-cylindrical rough element is exposed to an incoming turbulent shear flow, we would expect the spectra to be much more broadband than that noted for a cylinder exposed to a free-stream \citep{darekar2001}.
Indeed, on looking at the spectra for all the static cases, we found that both ${\phi}_{{p'}{p'}}$ and ${\phi}_{{C'_p}{C'_p}}$ are most energetic across a broadband frequency range, \rahul{$\mathcal{O}$(10$^{-3}$) $\lesssim$ $f{\nu}/{\overline{U}^{2}_{{\tau}_{s}}}$ $\lesssim$ $\mathcal{O}$(10$^{-2}$)} (shaded in magenta),
which is also exhibited by their cross-correlation spectra, ${\phi}_{{C'_p}{p'}}$ (not shown here to avoid clutter). 
This close correspondence in spectral energy distribution suggests that the pressure drag (which is represented by ${\phi}_{{C'_p}{C'_p}}$) is directly associated with the vortex shedding activity (represented by ${\phi}_{{p'}{p'}}$ for $\theta$ = $\pi$/2).
\rahulemp{It should be noted here that the correlation between ${\phi}_{{p'}{p'}}$ (at $\theta$ = $\pi$/2) and ${\phi}_{{C'_p}{C'_p}}$ is not obvious, given that the surface pressure at the crest does not contribute to the pressure drag along $x$.
Such a correlation is only possible if the pressure drag is dependant on the vortex shedding phenomena, which is established by comparing the two spectra.
Another noteworthy observation from figures \ref{fig5}(a,b) is that the energetic frequency range for both ${\phi}_{{p'}{p'}}$ and ${\phi}_{{C'_p}{C'_p}}$ remains consistent despite variation in the pressure drag domination across 6 $\le$ ${\Lambda}/k$ $\le$ 60.}
\rahuld{It suggests that despite the variation from fully to transitionally rough scenarios on increasing ${\Lambda}/k$, the pressure drag mechanism always corresponds to the same frequency range (for the present case of $k$ = 0.1$H$).}

Next, on considering wall-oscillations imposed at $T^+_{osc}$ $=$ 100 (figures \ref{fig5}c-e), the corresponding oscillation frequency lies within the most energetic frequency range of ${\phi}_{{p'}{p'}}$ and ${\phi}_{{C'_p}{C'_p}}$, \rahuld{for both $\Lambda$ = 10$k$ and 20$k$ corresponding to $k$ = 0.1$H$ and 0.2$H$ (see green arrows in figures \ref{fig5}c-e)}.
\rahulemp{One would thus expect the imposed wall oscillations to weaken the vortex shedding phenomena, likely by disturbing the two-dimensionality of the separated shear layer, and \emph{consequently} reducing the pressure drag \citep{darekar2001,choi2008}.
This explains the significant attenuation of ${\phi}_{{p'}{p'}}$ and ${\phi}_{{C'_p}{C'_p}}$, for the actuated cases ($T^+_{osc}$ = 100) as compared to the static cases. 
The same was also noted in case of their cross-correlation spectra, ${\phi}_{{C'_p}{p'}}$ (not shown here).
Hence, the choice of $T^+_{osc}$ = 100 lying within the energetic frequency range, for all rough wall cases (figures \ref{fig5}a,b), explains the significant pressure drag reduction noted for these cases.
This can be further confirmed by considering wall oscillations at a frequency much higher than the energetic range, for example $T^+_{osc}$ = 10 (figure \ref{fig5}c).
Indeed, both ${\phi}_{{p'}{p'}}$ and ${\phi}_{{C'_p}{C'_p}}$ spectra for $T^+_{osc}$ = 10 overlap with the corresponding spectra for the static case at $\Lambda$ = 10$k$, since such high oscillation frequency is incapable of disturbing the vortex shedding activity.} 
This explains the negligible DR\% corresponding to $T^+_{osc}$ = 10.
\rahuld{Along the same lines, in case of $k$ = 0.2$H$ (figure \ref{fig5}e), we can note that the attenuation of both the spectra (for actuated case relative to static case) is not as significant as noted for $k$ = 0.1$H$ (figure \ref{fig5}c), which again is consistent with the different magnitudes of DR\% for the two cases (figure \ref{fig2}b).
The results for $k$ = 0.2$H$, thus, support the notion that the magnitude of pressure drag reduction is correlated with the intensity with which shedding activity in the roughness wake is weakened (depicted by ${\phi}_{{p'}{p'}}$), and this is applicable despite the variation in roughness heights.
The results also confirm that imposition of wall oscillation frequency, of the order of the vortex shedding frequencies, leads to pressure drag reduction.}

\vspace{-3mm}
\section{Concluding remarks}
\label{future}

\rahulemp{The present study investigates drag reduction in a simplified (i.e. 2-D) rough wall scenario, with the intention to answer the following two open questions:
(i) Can the well-known viscous drag reduction strategy of imposing spanwise wall oscillations yield pressure drag reduction in a fully-rough case \rahul{(i.e. with negligible viscous drag contributions)}?
(ii) What is the mechanism driving this pressure drag reduction?}
These are answered here via a set of systematically conducted direct numerical simulations of a turbulent flow over 2-D semi-cylindrical rods, maintained at matched roughness Reynolds numbers ($k^+$). 
\rahuld{The study also considers a case for higher roughness Reynolds number, to investigate the applicability of the pressure drag reduction mechanism across different roughness heights}.
\rahulemp{It is found that spanwise wall oscillations are capable of attenuating the pressure drag when the wall oscillation frequency is of the order of frequency of vortex shedding from the rough elements. 
In all the rough wall cases simulated for the present study, the spectral distribution of the vortex shedding phenomena corresponded to a nominally similar range of viscous-scaled frequencies.}
This explained the pressure drag reduction on imposition of spanwise wall oscillations \rahul{for all rough wall cases, given they were imposed at constant viscous-scaled actuation parameters}.

\rahuld{Although the present study is limited to semi-cylindrical roughness geometry, the flow physics observed in association with the pressure drag reduction (namely, weakening of Reynolds stresses and increasing base pressure in the roughness wake) was found to be consistent with past observations noted in case of 2-D bumps \citep{banchetti2020} and square bar roughness \citep{nguyen2021}.
This supports applicability of the pressure drag reduction mechanism (i.e. weakening of vortex shedding) for varying roughness geometries, roughness heights and transitional/fully rough scenarios.
A thorough quantification of this argument, however, would require an extensive simulation campaign, which is beyond the scope of the present study.}

\rahulemp{A natural follow-up of this work would involve testing the drag reduction strategy on practical surface conditions, such as sand-paper or 3-D roughness geometries, and at high $Re_{\tau}$. 
In this case, the pressure drag is likely to be contributed from highly 3-D vortical structures shedding at very small viscous-scaled frequencies, thereby requiring significant input power to directly affect the vortex shedding (and associated pressure drag).}
In this respect, the ability to affect this high-frequency phenomenon, via low-frequency wall oscillations \citep{marusic2021}, by leveraging the inherent triadic interactions between the large- and small-time scales \citep{deshpande2022}, could offer an energy-efficient pathway for rough wall drag reduction.

\section*{\rahul{Appendix 1: Numerical parameters of the simulations}}
\label{sec:numerical-params}

\begin{table}
  \captionsetup{width=1.0\linewidth}
    \begin{tabular}{lccccccl}
      Case & $N_{\rm cyl}$ & $N_{\rm elem}$
      & $N_y$ & $\Delta x^+$& $\Delta y^+$
      & $\Delta z^+$ & $T_{{\rm obs}}\ubulk/\hmean$ (static/actuated)\\
      \hline
      $\Lambda = 6k$   & 10& 1340&192& 1.38--4.98& 4.7 & 0.62--7.29 & 418/713\\
      $\Lambda = 7.5k$ & 8 & 1296&192& 1.38--4.98& 4.7 &0.62--7.29&397/369\\
      $\Lambda = 10k$  & 6 & 1140&192& 1.38--4.98& 4.7 &0.62--7.29&401/747\\
      $\Lambda = 10k_{fine}$  & 6 & 2496&192& 0.60--3.3& 4.7 &0.28--6.29&\rahuld{268/187}\\
      \rahuld{$\Lambda = 10k^*$} & \rahuld{3} & \rahuld{1056}&\rahuld{192}& \rahuld{1.20--6.67}& \rahuld{4.7} &\rahuld{0.57--6.89}&\rahuld{194/181}\\
      $\Lambda = 12k$  & 5 & 950 &192& 1.38--4.98& 4.7 &0.63--7.29&412/383\\
      $\Lambda = 15k$  & 4 & 872 &192& 1.38--4.98& 4.7 &0.62--7.29&433/402\\
      $\Lambda = 20k$  & 3 & 822 &192& 1.38--4.98& 4.7 &0.62--7.29&471/1167\\
      $\Lambda = 60k$  & 1 & 666 &192& 1.38--4.98& 4.7 &0.62--7.29&489/586\\
      \hline
    \end{tabular}
    \caption{\rahul{Numerical paramters of the simulations. The computational box size 
      $L_x \times L_y \times L_z = 6\hmean\times 3\hmean \times \hmean$ is the same in all
      the simulations. 
      $N_{\rm cyl} = L_x/\cylspacing$ represents the number of roughness units 
      or streamwise periodic units in the computational box.
      $N_{\rm elem}$ is the number of two-dimensional spectral elements in the $x$-$z$
      plane, while $N_y$ is the number of Fourier modes in the homogeneous spanwise direction.
      $\Delta x^+$,  $\Delta y^+$ and $\Delta z^+$ are
      representative grid resolutions (normalized
      with the static viscous length scale, $\nu/{\overline{U}_{{\tau}_s}}$).
      $T_{\rm obs}$ is the steady-state simulation interval over which statistics are
      accumulated after discarding an initial transient.}
      \rahuld{Values marked with $^{*}$ correspond to the sole $k$ = 0.2$H$ case, while $k$ = 0.1$H$ for all other cases.}}
    \label{tab:numerical-parameters}
  \end{table}

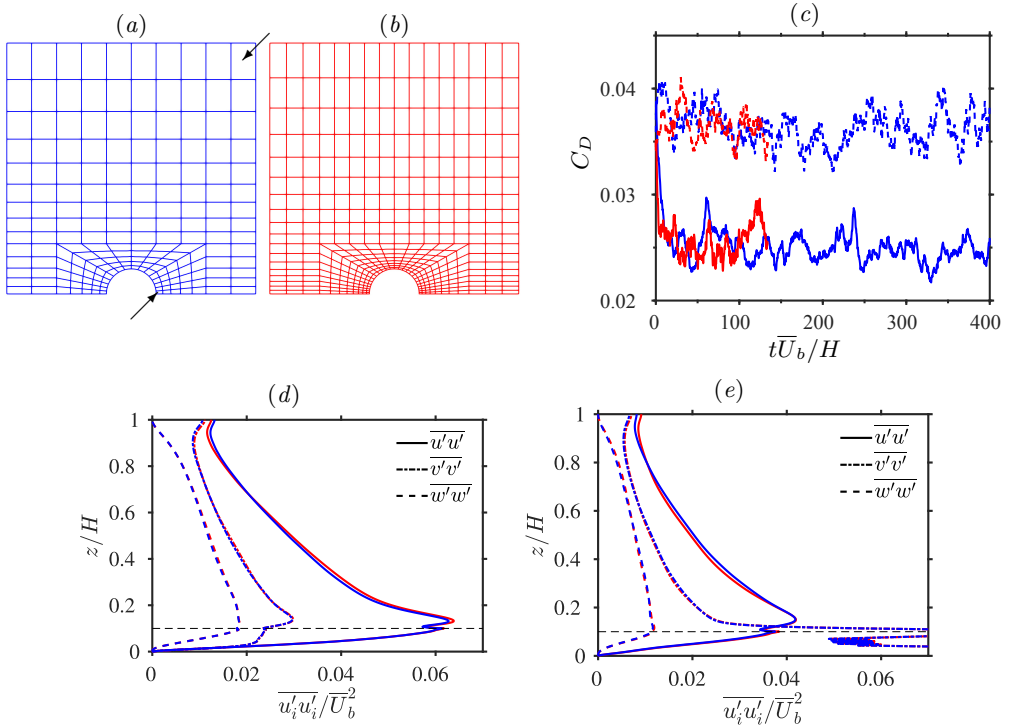
\begin{figure}
  \captionsetup{width=1.0\linewidth}
  \centering

  \begin{minipage}{0.52\linewidth}
    \begin{minipage}{0.48\linewidth}
      \centerline{(\textit{a})}
      \includegraphics[width=\linewidth]{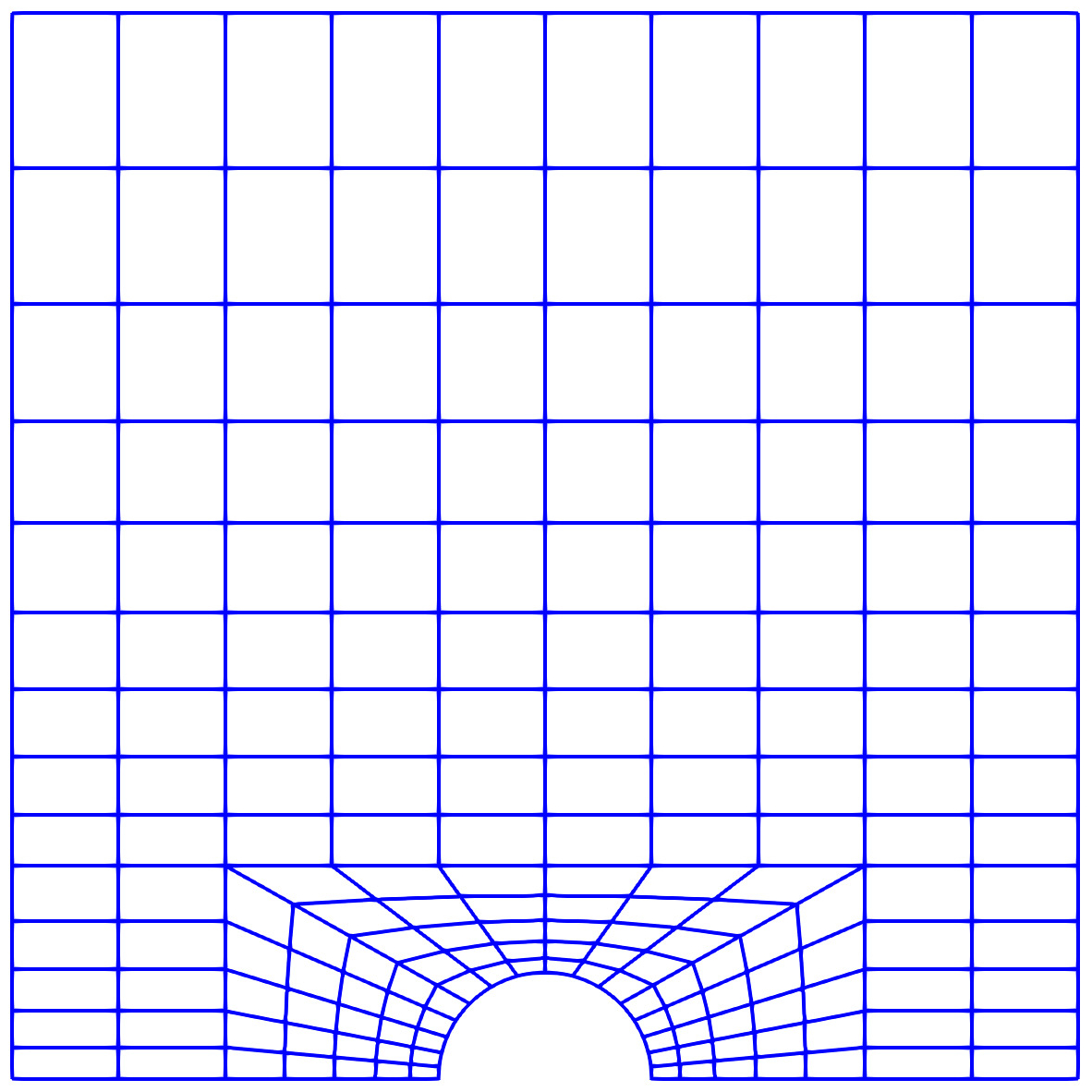}
    \end{minipage}
    \begin{minipage}{0.48\linewidth}
      \centerline{(\textit{b})}
      \includegraphics[width=\linewidth]{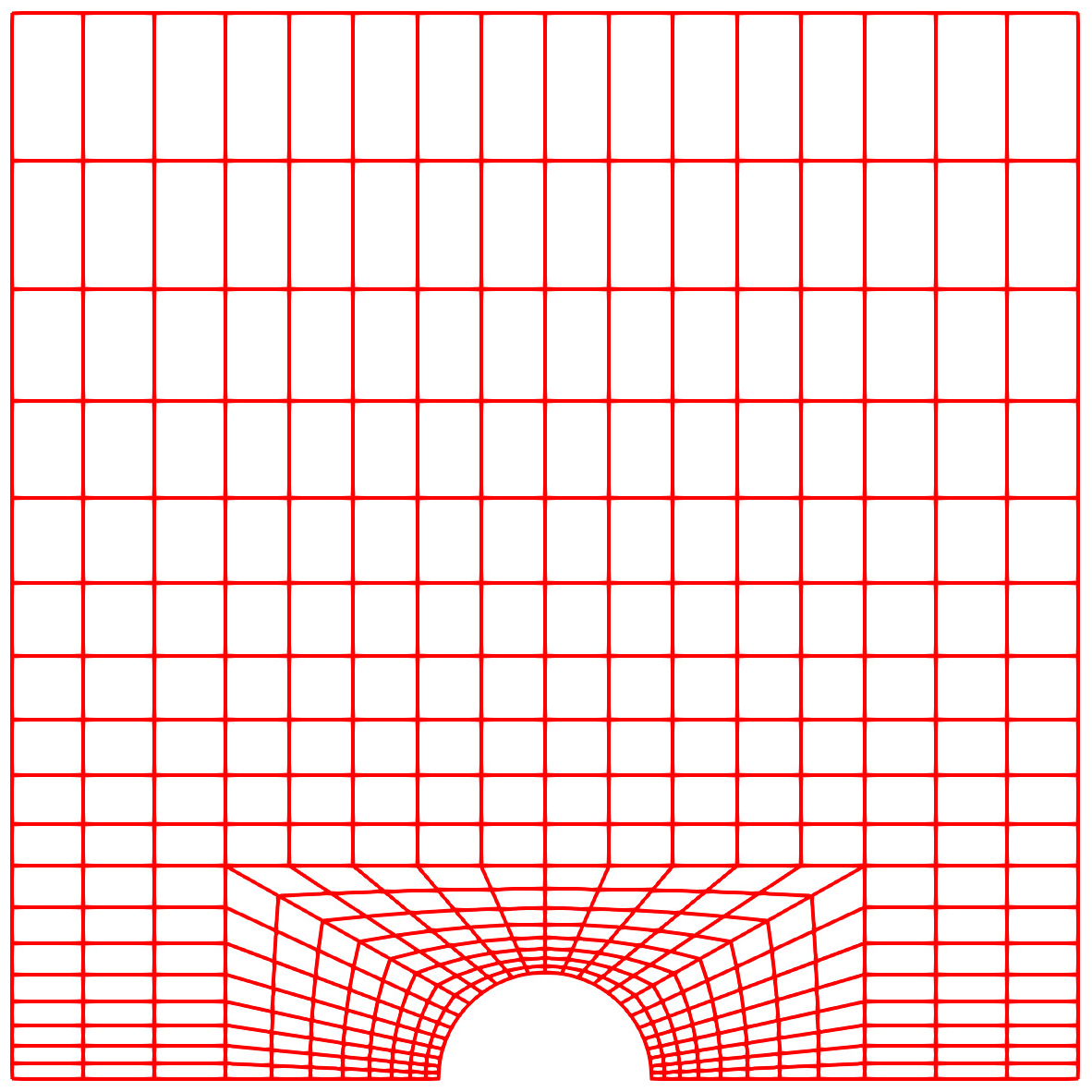}
    \end{minipage}%
    \begin{picture}(0,0)(0,0)
      \put(-147,-59){\vector(1,1){10}}%
      \put(-95,48){\vector(-1,-1){10}}%
    \end{picture}\\[20pt]
  \end{minipage}
  \hspace{2ex}
  \begin{minipage}{2ex}
    \rotatebox{90}{\hspace{6ex}$C_D$}
  \end{minipage}
  \begin{minipage}{0.4\linewidth}
    \centerline{(\textit{c})}
    \includegraphics[width=\linewidth]{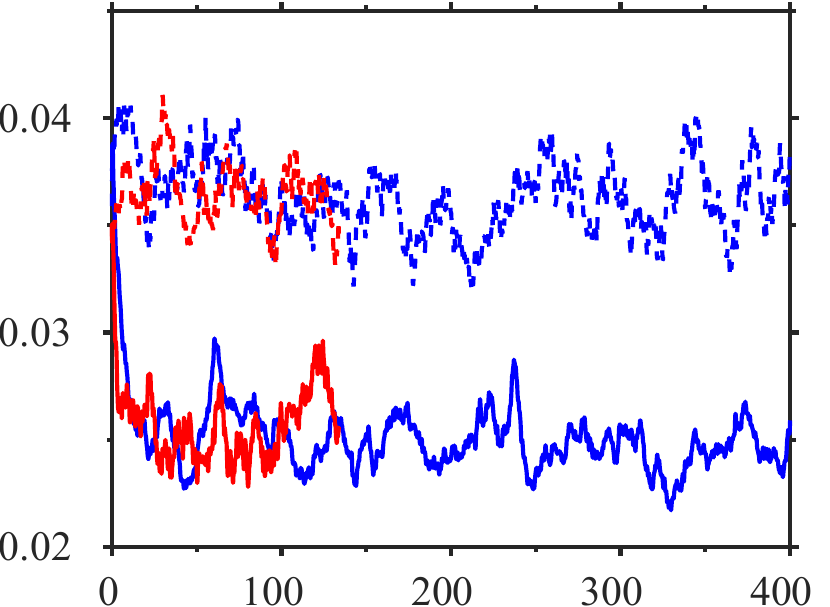}
    \centerline{$t \ubulk/\hmean$}
  \end{minipage}\\[5pt]
  \begin{minipage}{0.4\linewidth}
    \centerline{(\textit{d})}
    \includegraphics[width=\linewidth]{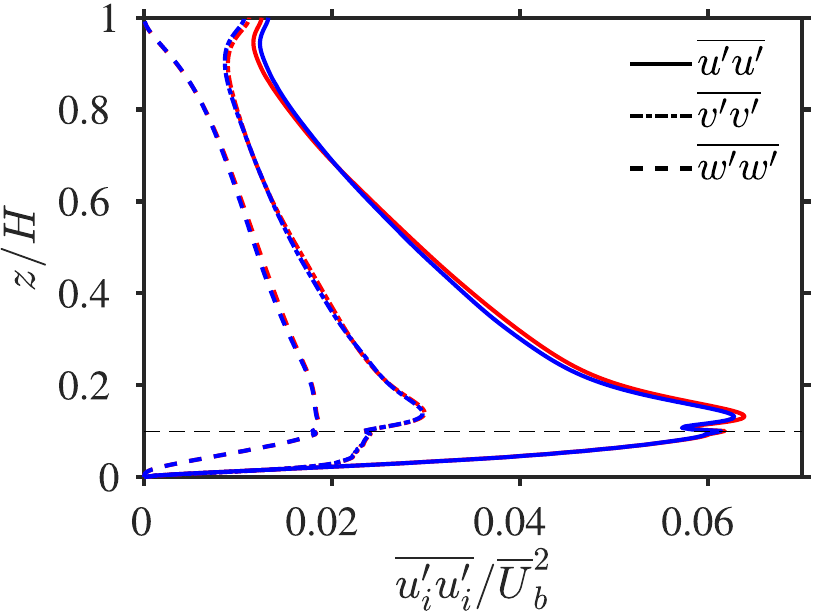}
  \end{minipage}
  \hspace{2ex}
  \begin{minipage}{0.4\linewidth}
    \centerline{(\textit{e})}
    \includegraphics[width=\linewidth]{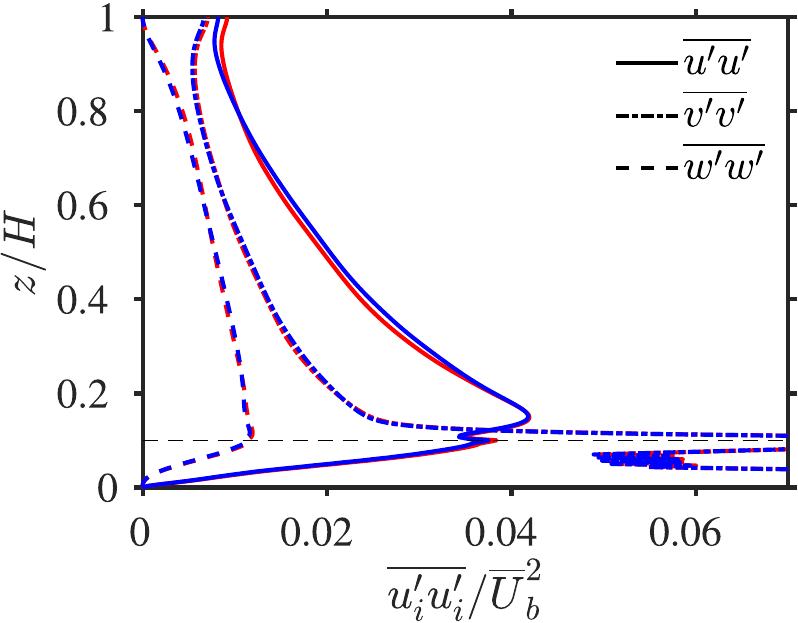}
  \end{minipage}
  \caption{\rahul{Depiction of the spectral element resolution in the 
    vicinity of the cylindrical roughness elements for (\textit{a}) case
    $\cylspacing=10\cylrad$ and (\textit{b}) case
    $\cylspacing=10\cylrad_{fine}$. The arrows point to the smallest and largest
    elements.
    (\textit{c}) Time evolution of the instantaneous total drag force coefficient
    for case $\cylspacing=10\cylrad$ (blue) and $\cylspacing=10\cylrad_{fine}$ (red).
    Solid lines show $C_D$ of the actuated case while the dashed lines show those of the
    corresponding static case.}
  \rahuld{(\textit{d,e}) Wall-normal profiles of the velocity fluctuation covariances
    (with respect to the global
  space and time averaged mean velocity $\overline{U}_i$) for the (d) static and (e) actuated cases corresponding to
  $\cylspacing=10\cylrad$ (blue) and $\cylspacing=10\cylrad_{fine}$ (red).}}

  \label{fig:grid-resolution-study}
 \end{figure}

\rahul{As discussed in \S\ref{setup}, the present simulations are carried out using
the open-source spectral element solver NEKTAR++ \citep{Cantwell2015}. In all
the production simulations performed, the grid resolution per the cylindrical roughness
element was kept identical. Each element was represented with
Lagrange polynomials of degree $N_p = 7$ on Gauss-Lobatto-Legendre points.
The computational parameters of the simulation cases are listed in table~\ref{tab:numerical-parameters}.
The size of the quadrilateral elements used to discretise the domain varies in space
with the smallest element located at the wall next to the cylinder edge
while the largest element is located at the top boundary
(shown by the arrows in figure~\ref{fig:grid-resolution-study}\textit{a}).
Denoting the streamwise and wall-normal size of the elements as $l^{elem}_x$ and
$l^{elem}_z$ respectively, the streamwise and wall-normal grid resolutions
within the smallest element can be estimated as
$\Delta x_{min}^+ = l^{elem+}_x/(N_p-1) ~\approx 1.38$ and
$\Delta z_{min}^+ = l^{elem+}_z/(N_p-1) ~\approx 0.624$. The corresponding grid resolution
within the largest element are $\Delta x^+_{max} \approx 4.98 $ and
$\Delta z^+_{max} \approx 7.29$
\rahuld{(as shown in table~\ref{tab:numerical-parameters}
  the values are slightly different for case $\Lambda = 10k^*$ in which $k=0.2H$)}.
The grid resolution in the homogeneous spanwise
direction is kept constant at $\Delta{y^+}$ $\approx$ 4.7.
These grid resolutions are more than sufficient to resolve the near-wall
turbulent scales of a smooth wall channel flow \citep{lozano2014}.
To ascertain the adequacy of the grid resolution to resolve the
turbulence associated with the vortex shedding downstream of the roughness elements,
we have run two additional simulations
(cases $\Lambda = 10k_{fine}$ in table \ref{tab:numerical-parameters})
that are identical to the static and actuated cases with $\Lambda = 10k$
but at a higher grid resolution (see figure~\ref{fig:grid-resolution-study}\textit{b}). 
In figure~\ref{fig:grid-resolution-study}(\textit{c}), we compare
the variation of the instantaneous total drag coefficient ($C_D$) as a function of time
for both the static and actuated cases at the two grid resolutions. The match between the temporal evolution of the $C_D$ values from both grid resolutions indicates the adequacy of the adopted grid resolution to capture the dominant pressure drag reduction mechanisms at the actuation frequencies of the present study.}
\rahuld{%
  Indeed, the mean drag coefficient values $\overline{C}_{D_{s}} = 36.1\times 10^{-3}$ and
  $\overline{C}_{D_{a}} = 25.0\times 10^{-3}$ for the static and actuated cases of   $\Lambda = 10k_{fine}$ are very close to
  the corresponding values of  $\Lambda = 10k$ (cf.\ table~\ref{tab1}).
  A more quantitative assessment of the grid resolution is provided in
  figures~\ref{fig:grid-resolution-study}(\textit{d,e}) that show the
  wall-normal profiles of the global velocity fluctuation covariances (averaged over the
  corresponding steady-state interval). Note that the
  fluctuations include the turbulent, form-induced and wall-oscillation contributions.
  The very good match observed between the profiles is a 
  further demonstration of the sufficiency of the considered resolution.
}

\begin{figure}
   \captionsetup{width=1.0\linewidth}
  \centering
   \begin{minipage}{3ex}
     \rotatebox{90}{\hspace{6ex}$k_x \phi_{u^\prime u^\prime}/\ufrics^2$}
   \end{minipage}
   \begin{minipage}{0.35\linewidth}
     \centerline{(\textit{a})}
     \includegraphics[width=\linewidth]{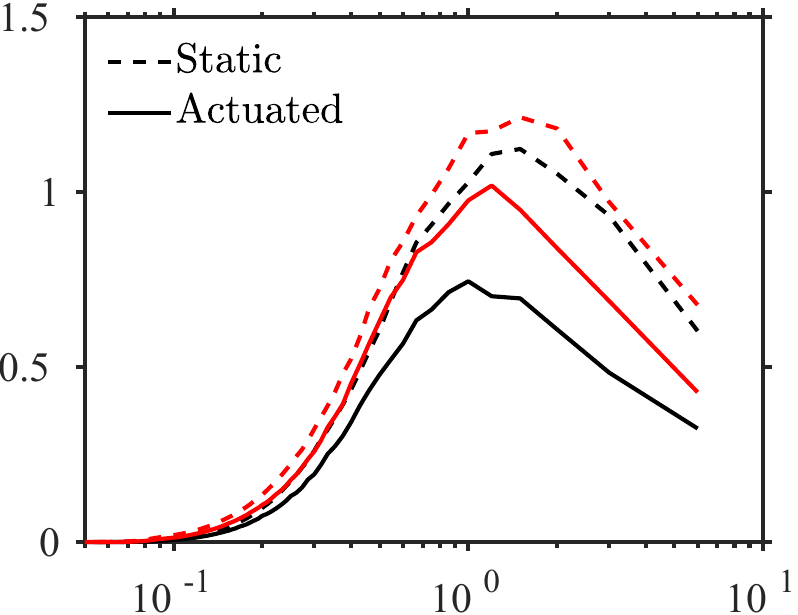}
     \centerline{$\lambda_x/\hmean$}
   \end{minipage}
   \hspace{1ex}
   \begin{minipage}{3ex}
     \rotatebox{90}{\hspace{6ex}$k_y \phi_{u^\prime u^\prime}/\ufrics^2$}
   \end{minipage}
   \begin{minipage}{0.35\linewidth}
     \centerline{(\textit{b})}
     \includegraphics[width=\linewidth]{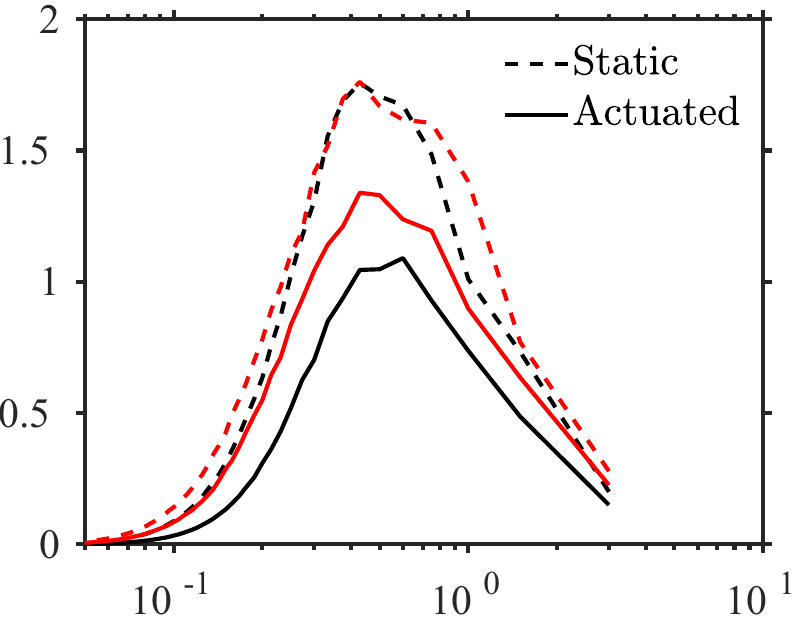}
     \centerline{$\lambda_y/\hmean$}
   \end{minipage}
   \caption{\rahul{(\textit{a}) Streamwise and (\textit{b}) spanwise pre-multiplied
     spectra of the streamwise velocity fluctuations on an $x$-$y$ plane located
     at $z^+\approx 35$. Black, $\cylspacing=10\cylrad$; red, $\cylspacing=60\cylrad$.}}
   \label{fig:plane-u-spectra}
 \end{figure}

 \rahul{In order to test the adequacy of adopted box size to accommodate the
 dominant turbulent structures, we have computed the streamwise and spanwise
 pre-multiplied energy spectra of the streamwise velocity fluctuation on an $x$--$y$
 plane located above the crest of the cylinders where the mean turbulent kinetic energy
 attains a maximum value (within the shear layer region).
 Figure \ref{fig:plane-u-spectra} shows the spectra for both the static and actuated
 cases of $\cylspacing=10\cylrad$ and $\cylspacing=60\cylrad$.
 Although very large box size is required to achieve complete de-correlation
 in the streamwise direction, it can be seen that the peak of the energy spectra,
 both in the streamwise and spanwise directions, are captured
 in the box size $L_x \times L_y \times L_z = 6\hmean\times 3\hmean \times \hmean$.}

\vspace{-4mm}
\section*{Acknowledgments}
Financial support from the Australian Research Council (Grant number: DP210102172) and the Office of Naval Research (Grant number: N62909-23-1-2068) is gratefully acknowledged. 
R. Deshpande also acknowledges partial financial support by the University of Melbourne through the Melbourne Postdoctoral Fellowship.
This research was undertaken with the assistance of resources from the National Computational Infrastructure (NCI Australia), an NCRIS enabled capability supported by the Australian Government.

\vspace{-4mm}
\section*{Declaration of Interests} 

The authors report no conflict of interest.

\bibliographystyle{jfm}
\bibliography{DragReduction_2dRoughWallOsc_bib}

\begin{thebibliography}{47}
\expandafter\ifx\csname natexlab\endcsname\relax\def\natexlab#1{#1}\fi
\def\au#1{#1} \def\ed#1{#1} \def\yr#1{#1}\def\at#1{#1}\def\jt#1{\textit{#1}} \def\bt#1{#1}\def\bvol#1{\textbf{#1}} \def\vol#1{#1} \def\pg#1{#1} \def\publ#1{#1}\def\arxiv#1{#1}\def\org#1{#1}\def\st#1{\textit{#1}}

\bibitem[Akhavan {\em et~al.\/}(1993)Akhavan, Jung \& Mangiavacchi]{akhavan1993}
{\sc \au{Akhavan, R.}, \au{Jung, W.} \& \au{Mangiavacchi, N.}} \yr{1993}  \at{Control of wall turbulence by high frequency spanwise oscillations}.  \jt{AIAA}  \bvol{93},  \pg{3282}.

\bibitem[Alam(2022)]{alam2022}
{\sc \au{Alam, M.~M.}} \yr{2022}  \at{A review of cylinder corner effect on flow and heat transfer}.  \jt{Journal of Wind Engineering and Industrial Aerodynamics}  \bvol{229},  \pg{105132}.

\bibitem[Banchetti {\em et~al.\/}(2020)Banchetti, Luchini \& Quadrio]{banchetti2020}
{\sc \au{Banchetti, J.}, \au{Luchini, P.} \& \au{Quadrio, M.}} \yr{2020}  \at{Turbulent drag reduction over curved walls}.  \jt{J. Fluid Mech.}  \bvol{896},  \pg{A10}.

\bibitem[Bandyopadhyay(1986)]{bandyopadhyay1986}
{\sc \au{Bandyopadhyay, P.~R.}} \yr{1986}  \at{Drag reducing outer-layer devices in rough wall turbulent boundary layers}.  \jt{Exp. Fluids}  \bvol{4}~(5),  \pg{247--256}.

\bibitem[Bearman \& Owen(1998)]{bearman1998}
{\sc \au{Bearman, P.~W.} \& \au{Owen, J.~C.}} \yr{1998}  \at{Reduction of bluff-body drag and suppression of vortex shedding by the introduction of wavy separation lines}.  \jt{J. Fluids Struct.}  \bvol{12}~(1),  \pg{123--130}.

\bibitem[Brackston {\em et~al.\/}(2016)Brackston, Wynn \& Morrison]{brackston2016}
{\sc \au{Brackston, R.~D.}, \au{Wynn, A.} \& \au{Morrison, J.~F.}} \yr{2016}  \at{Extremum seeking to control the amplitude and frequency of a pulsed jet for bluff body drag reduction}.  \jt{Exp. Fluids}  \bvol{57}~(10),  \pg{1--14}.

\bibitem[Cantwell {\em et~al.\/}(2015)Cantwell, Moxey, Comerford, Bolis {\em et~al.\/}]{Cantwell2015}
{\sc \au{Cantwell, C.D.}, \au{Moxey, D.}, \au{Comerford, A.}, \au{Bolis, A.} \& \au{others}} \yr{2015}  \at{Nektar++: An open-source spectral/hp element framework}.  \jt{Comput. Phys. Commun.}  \bvol{192},  \pg{205--219}.

\bibitem[Chan {\em et~al.\/}(2015)Chan, MacDonald, Chung, Hutchins \& Ooi]{chan2015}
{\sc \au{Chan, L.}, \au{MacDonald, M.}, \au{Chung, D.}, \au{Hutchins, N.} \& \au{Ooi, A.}} \yr{2015}  \at{A systematic investigation of roughness height and wavelength in turbulent pipe flow in the transitionally rough regime}.  \jt{J. Fluid Mech.}  \bvol{771},  \pg{743--777}.

\bibitem[Cho {\em et~al.\/}(2016)Cho, Choi \& Choi]{cho2016}
{\sc \au{Cho, M.}, \au{Choi, S.} \& \au{Choi, H.}} \yr{2016}  \at{Control of flow separation in a turbulent boundary layer using time-periodic forcing}.  \jt{J. Fluids Eng.}  \bvol{138}~(10).

\bibitem[Choi {\em et~al.\/}(2008)Choi, Jeon \& Kim]{choi2008}
{\sc \au{Choi, H.}, \au{Jeon, W.~P.} \& \au{Kim, J.}} \yr{2008}  \at{Control of flow over a bluff body}.  \jt{Annu. Rev. Fluid Mech.}  \bvol{40},  \pg{113--139}.

\bibitem[Chopra \& Mittal(2022)]{chopra2022}
{\sc \au{Chopra, G.} \& \au{Mittal, S.}} \yr{2022}  \at{Secondary vortex, laminar separation bubble and vortex shedding in flow past a low aspect ratio circular cylinder}.  \jt{J. Fluid Mech.}  \bvol{930},  \pg{A12}.

\bibitem[Chung {\em et~al.\/}(2021)Chung, Hutchins, Schultz \& Flack]{chung2021}
{\sc \au{Chung, D.}, \au{Hutchins, N.}, \au{Schultz, M.~P.} \& \au{Flack, K.~A.}} \yr{2021}  \at{Predicting the drag of rough surfaces}.  \jt{Annu. Rev. Fluid Mech.}  \bvol{53},  \pg{439--471}.

\bibitem[Coceal {\em et~al.\/}(2007)Coceal, Dobre, Thomas \& Belcher]{coceal2007}
{\sc \au{Coceal, O.}, \au{Dobre, A.}, \au{Thomas, T.~G.} \& \au{Belcher, S.~E.}} \yr{2007}  \at{Structure of turbulent flow over regular arrays of cubical roughness}.  \jt{J. Fluid Mech.}  \bvol{589},  \pg{375--409}.

\bibitem[Corke {\em et~al.\/}(1982)Corke, Nagib \& Guezennec]{corke1982}
{\sc \au{Corke, T.~C.}, \au{Nagib, H.~M.} \& \au{Guezennec, Y.~G.}} \yr{1982}  \bt{A new view on origin, role and manipulation of large scales in turbulent boundary layers}.  \org{{\em Tech. Rep.\/}}.

\bibitem[Corke \& Thomas(2018)]{corke2018}
{\sc \au{Corke, T.~C.} \& \au{Thomas, F.~O.}} \yr{2018}  \at{Active and passive turbulent boundary-layer drag reduction}.  \jt{AIAA}  \bvol{56}~(10),  \pg{3835--3847}.

\bibitem[Darekar \& Sherwin(2001)]{darekar2001}
{\sc \au{Darekar, R.~M.} \& \au{Sherwin, S.~J.}} \yr{2001}  \at{Flow past a square-section cylinder with a wavy stagnation face}.  \jt{J. Fluid Mech.}  \bvol{426},  \pg{263--295}.

\bibitem[Desai {\em et~al.\/}(2020)Desai, Mittal \& Mittal]{desai2020}
{\sc \au{Desai, A.}, \au{Mittal, S.} \& \au{Mittal, S.}} \yr{2020}  \at{Experimental investigation of vortex shedding past a circular cylinder in the high subcritical regime}.  \jt{Phys. Fluids}  \bvol{32}~(1),  \pg{014105}.

\bibitem[Deshpande {\em et~al.\/}(2023)Deshpande, Chandran, Smits \& Marusic]{deshpande2022}
{\sc \au{Deshpande, R.}, \au{Chandran, D.}, \au{Smits, A.~J.} \& \au{Marusic, I.}} \yr{2023}  \at{On the relationship between manipulated inter-scale phase and energy-efficient turbulent drag reduction}.  \jt{Journal of Fluid Mechanics}  \bvol{972},  \pg{A12}.

\bibitem[Deshpande {\em et~al.\/}(2017)Deshpande, Desai, Kanti \& Mittal]{deshpande2017}
{\sc \au{Deshpande, R.}, \au{Desai, A.}, \au{Kanti, V.} \& \au{Mittal, S.}} \yr{2017} Experimental investigation of boundary layer transition in flow past a bluff body.  \bt{In {\em Journal of Physics: Conference Series\/}}, ,  \vol{vol. 822},  \pg{p. 012003}. IOP Publishing.

\bibitem[Furuya {\em et~al.\/}(1976)Furuya, Miyata \& Fujita]{furuya1976}
{\sc \au{Furuya, Y.}, \au{Miyata, M.} \& \au{Fujita, H.}} \yr{1976}  \at{Turbulent boundary layer and flow resistance on plates roughened by wires}.  \jt{Trans. ASME: J. Fluids Eng.}  \bvol{98},  \pg{635--644}.

\bibitem[Garcia {\em et~al.\/}(2021)Garcia, Ahmad \& Hussain]{garcia2021}
{\sc \au{Garcia, E.}, \au{Ahmad, A.} \& \au{Hussain, F.}} \yr{2021} Drag control on a fully rough oscillating flat plate.  \bt{In {\em APS Division of Fluid Dynamics Meeting Abstracts\/}},  \pg{pp. M23--001}.

\bibitem[Gatti \& Quadrio(2013)]{gatti2013}
{\sc \au{Gatti, D.} \& \au{Quadrio, M.}} \yr{2013}  \at{Performance losses of drag-reducing spanwise forcing at moderate values of the {R}eynolds number}.  \jt{Physics of Fluids}  \bvol{25}~(12).

\bibitem[Gatti \& Quadrio(2016)]{gatti2016}
{\sc \au{Gatti, D.} \& \au{Quadrio, M.}} \yr{2016}  \at{Reynolds-number dependence of turbulent skin-friction drag reduction induced by spanwise forcing}.  \jt{J. Fluid Mech.}  \bvol{802},  \pg{553--582}.

\bibitem[Hwang {\em et~al.\/}(2013)Hwang, Kim \& Choi]{hwang2013}
{\sc \au{Hwang, Y.}, \au{Kim, J.} \& \au{Choi, H.}} \yr{2013}  \at{Stabilization of absolute instability in spanwise wavy two-dimensional wakes}.  \jt{Journal of Fluid Mechanics}  \bvol{727},  \pg{346--378}.

\bibitem[Jim{\'e}nez(2004)]{jimenez2004}
{\sc \au{Jim{\'e}nez, J.}} \yr{2004}  \at{Turbulent flows over rough walls}.  \jt{Annu. Rev. Fluid Mech.}  \bvol{36},  \pg{173--196}.

\bibitem[Jim{\'e}nez(2018)]{jimenez2018}
{\sc \au{Jim{\'e}nez, J.}} \yr{2018}  \at{Coherent structures in wall-bounded turbulence}.  \jt{J. Fluid Mech.}  \bvol{842}.

\bibitem[Kim(2011)]{kim2011}
{\sc \au{Kim, J.}} \yr{2011}  \at{Physics and control of wall turbulence for drag reduction}.  \jt{Phil. Trans. R. Soc. A}  \bvol{369}~(1940),  \pg{1396--1411}.

\bibitem[Kim \& Choi(2005)]{kim2005}
{\sc \au{Kim, J.} \& \au{Choi, H.}} \yr{2005}  \at{Distributed forcing of flow over a circular cylinder}.  \jt{Phys. Fluids}  \bvol{17}~(3),  \pg{033103}.

\bibitem[Lee {\em et~al.\/}(2011)Lee, Sung \& Krogstad]{lee2011cube}
{\sc \au{Lee, J.~H.}, \au{Sung, H.~J.} \& \au{Krogstad, P.}} \yr{2011}  \at{Direct numerical simulation of the turbulent boundary layer over a cube-roughened wall}.  \jt{J. Fluid Mech.}  \bvol{669},  \pg{397--431}.

\bibitem[Lee \& Sung(2007)]{lee2007}
{\sc \au{Lee, S.} \& \au{Sung, H.~J.}} \yr{2007}  \at{Direct numerical simulation of the turbulent boundary layer over a rod-roughened wall}.  \jt{J. Fluid Mech.}  \bvol{584},  \pg{125--146}.

\bibitem[Leonardi {\em et~al.\/}(2015)Leonardi, Orlandi, Djenidi \& Antonia]{leonardi2015}
{\sc \au{Leonardi, S.}, \au{Orlandi, P.}, \au{Djenidi, L.} \& \au{Antonia, R.A.}} \yr{2015}  \at{Heat transfer in a turbulent channel flow with square bars or circular rods on one wall}.  \jt{J. Fluid Mech.}  \bvol{776},  \pg{512--530}.

\bibitem[Leonardi {\em et~al.\/}(2003)Leonardi, Orlandi {\em et~al.\/}]{leonardi2003}
{\sc \au{Leonardi, S.}, \au{Orlandi, P.} \& \au{others}} \yr{2003}  \at{Direct numerical simulations of turbulent channel flow with transverse square bars on one wall}.  \jt{J. Fluid Mech.}  \bvol{491},  \pg{229--238}.

\bibitem[Lin {\em et~al.\/}(1990)Lin, Howard \& Selby]{lin1990}
{\sc \au{Lin, J.~C.}, \au{Howard, F.~G.} \& \au{Selby, G.~V.}} \yr{1990}  \at{Small submerged vortex generators for turbulent flow separation control}.  \jt{J. Spacecr. Rockets.}  \bvol{27}~(5),  \pg{503--507}.

\bibitem[Lozano-Dur{\'a}n \& Jim{\'e}nez(2014)]{lozano2014}
{\sc \au{Lozano-Dur{\'a}n, A.} \& \au{Jim{\'e}nez, J.}} \yr{2014}  \at{Effect of the computational domain on direct simulations of turbulent channels up to ${R}e_{\tau}$ = 4200}.  \jt{Physics of Fluids}  \bvol{26}~(1).

\bibitem[Marusic {\em et~al.\/}(2021)Marusic, Chandran, Rouhi, Fu, Wine, Holloway, Chung \& Smits]{marusic2021}
{\sc \au{Marusic, I.}, \au{Chandran, D.}, \au{Rouhi, A.}, \au{Fu, M.~K.}, \au{Wine, D.}, \au{Holloway, B.}, \au{Chung, D.} \& \au{Smits, A.~J.}} \yr{2021}  \at{An energy-efficient pathway to turbulent drag reduction}.  \jt{Nat. Comm.}  \bvol{12}~(1),  \pg{1--8}.

\bibitem[Miyake {\em et~al.\/}(2001)Miyake, Tsujimoto \& Nakaji]{miyake2001}
{\sc \au{Miyake, Y.}, \au{Tsujimoto, K.} \& \au{Nakaji, M.}} \yr{2001}  \at{Direct numerical simulation of rough-wall heat transfer in a turbulent channel flow}.  \jt{Int. J. Heat and Fluid Flow}  \bvol{22}~(3),  \pg{237--244}.

\bibitem[Nagano {\em et~al.\/}(2004)Nagano, Hattori \& Houra]{nagano2004}
{\sc \au{Nagano, Y.}, \au{Hattori, H.} \& \au{Houra, T.}} \yr{2004}  \at{{DNS} of velocity and thermal fields in turbulent channel flow with transverse-rib roughness}.  \jt{Int. J. Heat and Fluid Flow}  \bvol{25}~(3),  \pg{393--403}.

\bibitem[Nguyen {\em et~al.\/}(2021)Nguyen, Ricco \& Pironti]{nguyen2021}
{\sc \au{Nguyen, V.}, \au{Ricco, P.} \& \au{Pironti, G.}} \yr{2021}  \at{Separation drag reduction through a spanwise oscillating pressure gradient}.  \jt{J. Fluid Mech.}  \bvol{912},  \pg{A20}.

\bibitem[Owen {\em et~al.\/}(2001)Owen, Bearman \& Szewczyk]{owen2001}
{\sc \au{Owen, J.~C.}, \au{Bearman, P.~W.} \& \au{Szewczyk, A.~A.}} \yr{2001}  \at{Passive control of {VIV} with drag reduction}.  \jt{J. Fluids Struct.}  \bvol{15}~(3-4),  \pg{597--605}.

\bibitem[Post \& Corke(2006)]{post2006}
{\sc \au{Post, M.~L.} \& \au{Corke, T.~C.}} \yr{2006}  \at{Separation control using plasma actuators: dynamic stall vortex control on oscillating airfoil}.  \jt{AIAA}  \bvol{44}~(12),  \pg{3125--3135}.

\bibitem[Quadrio {\em et~al.\/}(2022)Quadrio, Chiarini, Banchetti, Gatti, Memmolo \& Pirozzoli]{quadrio2022}
{\sc \au{Quadrio, M.}, \au{Chiarini, A.}, \au{Banchetti, J.}, \au{Gatti, D.}, \au{Memmolo, A.} \& \au{Pirozzoli, S.}} \yr{2022}  \at{Drag reduction on a transonic airfoil}.  \jt{Journal of Fluid Mechanics}  \bvol{942},  \pg{R2}.

\bibitem[Quadrio {\em et~al.\/}(2009)Quadrio, Ricco \& Viotti]{quadrio2009}
{\sc \au{Quadrio, M.}, \au{Ricco, P.} \& \au{Viotti, C.}} \yr{2009}  \at{Streamwise-travelling waves of spanwise wall velocity for turbulent drag reduction}.  \jt{J. Fluid Mech.}  \bvol{627},  \pg{161--178}.

\bibitem[Ricco {\em et~al.\/}(2021)Ricco, Skote \& Leschziner]{ricco2021}
{\sc \au{Ricco, P.}, \au{Skote, M.} \& \au{Leschziner, M.~A.}} \yr{2021}  \at{A review of turbulent skin-friction drag reduction by near-wall transverse forcing}.  \jt{Prog. Aero. Sci.}  \bvol{123},  \pg{100713}.

\bibitem[Seifert \& Pack(2002)]{seifert2002}
{\sc \au{Seifert, A.} \& \au{Pack, L.~G.}} \yr{2002}  \at{Active flow separation control on wall-mounted hump at high {R}eynolds numbers}.  \jt{AIAA}  \bvol{40}~(7),  \pg{1363--1372}.

\bibitem[Son {\em et~al.\/}(2011)Son, Choi, Jeon \& Choi]{son2011}
{\sc \au{Son, K.}, \au{Choi, J.}, \au{Jeon, W.P.} \& \au{Choi, H.}} \yr{2011}  \at{Mechanism of drag reduction by a surface trip wire on a sphere}.  \jt{Journal of Fluid Mechanics}  \bvol{672},  \pg{411--427}.

\bibitem[Wu {\em et~al.\/}(2020)Wu, Christensen \& Pantano]{wu2020}
{\sc \au{Wu, S.}, \au{Christensen, K.~T.} \& \au{Pantano, C.}} \yr{2020}  \at{A study of wall shear stress in turbulent channel flow with hemispherical roughness}.  \jt{J. Fluid Mech.}  \bvol{885},  \pg{A16}.

\bibitem[Yakeno {\em et~al.\/}(2015)Yakeno, Kawai, Nonomura \& Fujii]{yakeno2015}
{\sc \au{Yakeno, A.}, \au{Kawai, S.}, \au{Nonomura, T.} \& \au{Fujii, K.}} \yr{2015}  \at{Separation control based on turbulence transition around a two-dimensional hump at different {R}eynolds numbers}.  \jt{Int. J. Heat and Fluid Flow}  \bvol{55},  \pg{52--64}.

\end{thebibliography}

\end{document}